\documentclass[usenatbib]{mn2e}
\usepackage{times}
\usepackage{graphicx}
\usepackage[ansinew]{inputenc}
\usepackage[T1]{fontenc}
\usepackage{lmodern}
\usepackage{amssymb,amsmath}
\usepackage{rotating}
\usepackage{multirow}
\usepackage{pdflscape}
\numberwithin{equation}{section}

\title[Two estimates of $ R_0 $]{Two estimates of the distance to the Galactic Centre}
\author[Charles Francis and Erik Anderson]{Charles Francis$^{1}$\thanks{E-mail: C.E.H.Francis.75@cantab.net} and Erik Anderson$^{2}$\\
$^{1}$Jesus College, University of Cambridge, Cambridge CB5 8BL, UK.\\
$^{2}$360 Iowa Street, Ashland, OR 97520, USA}

\begin{document}

\pagerange{\pageref{firstpage}--\pageref{lastpage}} \pubyear{2014}

\maketitle

\label{firstpage}

\begin{abstract}
We use recently updated globular cluster distances to estimate the distance to the Galactic Centre, finding $ 7.4 \pm 0.2|_{\mathrm{stat}} \pm 0.2|_{\mathrm{sys}} $ kpc from symmetry considerations, including a trough at the Galactic Centre and peaks denoting the position of the bar. We recalibrate the red clump magnitude from Hipparcos stars, finding a skew distribution and a significant difference between peak and mean magnitudes. We find an estimate from stars in the periphery of the bulge using 2MASS, $ R_0 = 7.5 \pm 0.3 $ kpc, in agreement with the figure from the halo centroid. We resolve discrepancies in the literature between estimates from the red clump. Our results are consistent with those found by different methodologies after taking systematic errors into account. 
\end{abstract}


\section{Introduction}\label{sec:1}
As described by, e.g., Reid (1993), the distance, $ R_0 $ kpc, to the Galactic Centre is one of the most important parameters in Galactic astronomy, with implications ranging from Galactic dynamics and luminosity to extragalactic distance scales and the value of Hubble's constant. Extensive efforts have been made for nearly a century to determine its value accurately, but even in recent years agreement between measurements has not been reached, and Reid and Brunthaler's (2004) determination of the proper motion of Sgr A* has shown that the IAU standard value (Kerr \& Lynden-Bell 1986) $ R_0 = 8.5 $ kpc is not compatible with the IAU standard value of the rate of Galactic rotation, $ \Theta_0 = 220 $ km~s$^{-1}$.

In our view, the determination of the halo centroid is one of the more robust methods of determination of $ R_0 $ kpc, because symmetries seen in the structure of the distribution are independent of population incompleteness, distances of the majority of globular clusters are accurately determined from a number of independent measurements and little affected by reddening, and because statistical properties of a population are invariably more precisely determined than properties of individual members. The most recent published study using this method, Bica et al. (2006, hereafter B06), found $ R_0 = 7.2 \pm 0.3 $ kpc, which is significantly less than a number of recent determinations using other methods. Cluster distances have already been revised upwards, following the influential recalibration by Reid (1997) of the RR Lyrae scale from distances to subdwarfs in the Hipparcos catalogue (Perryman et al. 1997). It is appropriate therefore to re-examine the distribution of globular clusters, taking advantage of recent measurements. If the difference between this estimate and others is not resolved it will be necessary to consider whether the RR Lyrae scale requires further revision, or whether other estimates of $R_0$ are at fault.

To study this question, we recalculated distances for 154 clusters in the McMaster catalogue (Harris 1996, 2010 edition, hereafter H10) from a total of 560 recent measurements of distance (section 2; see online catalogue). H10 gives recalibrated distances of 157 probable globular clusters, based, in most cases, on single studies of the HB magnitude taken from the literature. We have found mean distances from multiple studies including isochrones, main sequence fitting, dynamical parallax, eclipsing binary and measurements of $K$ mag, and whenever the appropriate magnitude and reddening is determined in the source, we have used distances calibrated to the $ M_V(\mathrm{RR}) - \mathrm{[M/H]} $ relation of Catelan et al (2004), with corrections to $ M_V(\mathrm{ZAHB})$ and $ M_V(\mathrm{HB})$ as suggested by Sandage (1993) and Caloi (1997), together with metallicities updated by Saviane et al. (2012). We separately used H10 and our catalogue to estimate $R_0$ (section 3), finding $ 7.4 \pm 0.2|_{\mathrm{stat}} \pm 0.2|_{\mathrm{sys}} $ kpc, only a little greater than B06. 

In section 4 we recalibrate the red clump magnitude from nearby Hipparcos stars with parallaxes from the Hipparcos New Reduction (van Leeuwen 2007, hereinafter HNR). In section 5 we recalculated $R_0$ from stars in the periphery of the bulge in 2MASS. The use of the $K$ band and the relatively high latitudes of these stars minimises errors due to reddening. We found $ R_0 = 7.5 \pm 0.3 $ kpc in good agreement with the value from the halo centroid. Until results from Gaia become available, the HNR remains the primary source of accurate parallaxes in the solar neighbourhood. Its accuracy has been questioned by some astronomers, because of the low parallax distance to the Pleiades found by van Leeuwen (2009), but Francis \& Anderson (2012b) showed that this is because correlations between parallax and parallax error for cluster stars invalidate the weighted mean used by van Leeuwen. When the straight mean is used there are no systematic or anomalous differences between cluster distances measured by HNR parallax and those using other methods.

Our results are toward the lower end of the range of estimates of the distance to the Galactic Centre (section 6). We compared them with eight other estimates from the red clump using the $K$ and $I$ bands, which are not all in agreement. We trace apparent conflict between these estimates to factors such as the difference between the mean and peak red clump magnitude, selection biases in samples used for calibration (as previously reported by Groenewegen 2008) and the question of non-standard extinction toward the Galactic Centre. We have found good agreement between our estimates and results from circular motion tracers and the motion of S2 ignoring problematic data when S2 was close to pericentre. 

\section{Cluster distances}\label{Cluster distances}
We have calculated mean distances from a total of 560 measurements of distance from the recent literature for 154 globular cluster out of 157 given in the McMaster catalogue (H10). Our database includes estimates from a full range of methods, including main sequence fitting, the red clump magnitude, dynamical parallax, eclipsing binary and measurements of horizontal branch stars in the $K$ mag, but the greatest number use $ V $ magnitudes of the horizontal branch. We elected to determine distances for these measurements using updated figures for metallicity and a homogeneous calibration of the $ M_V(\mathrm{RR}) - \mathrm{[Fe/H]} $ relation. For other measurements we used the estimate of $R_0$ kpc given by the source.
\begin{figure}
	\centering
		\includegraphics[width=0.47\textwidth]{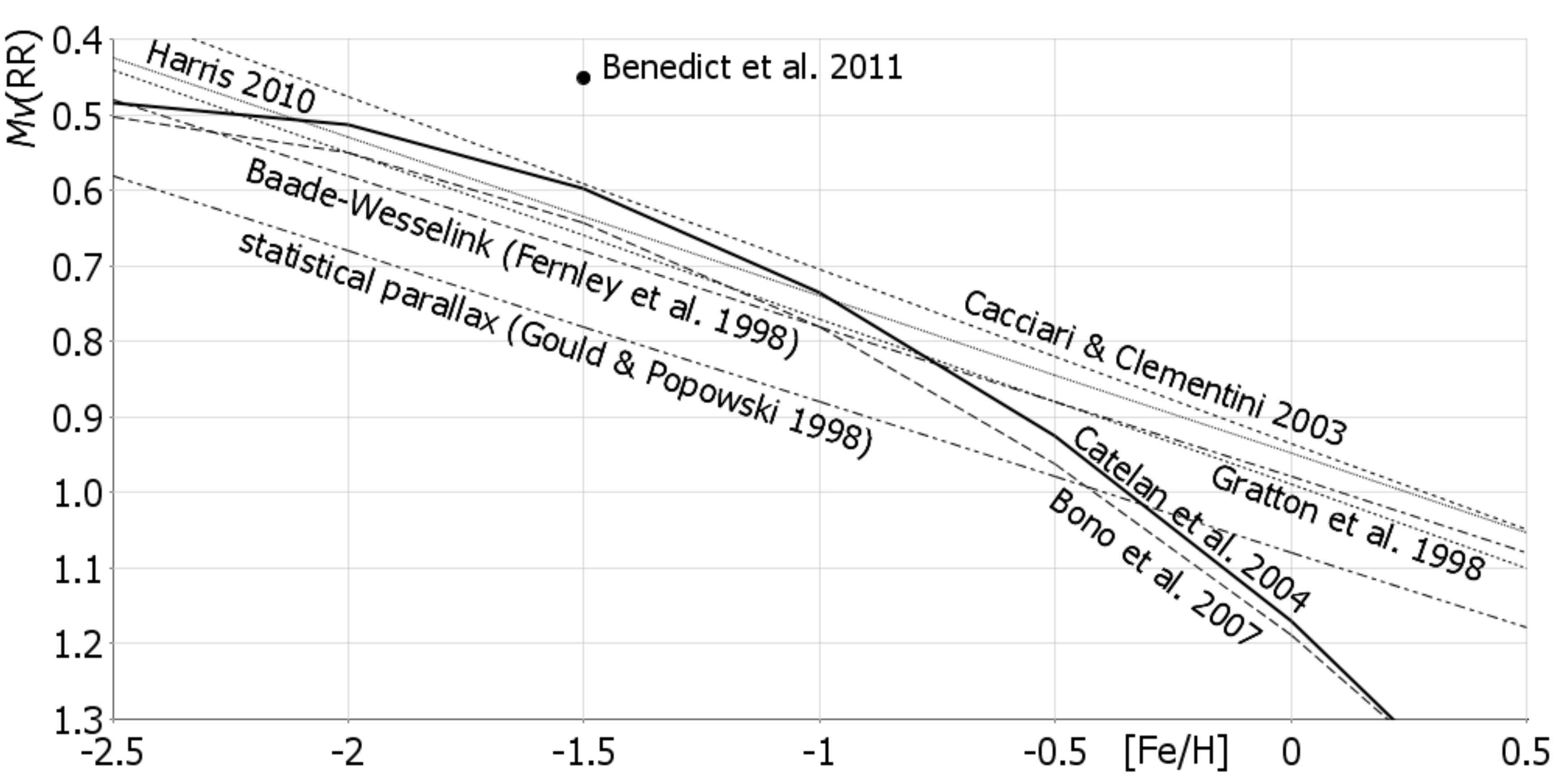}
	\caption{The dependency of RR Lyrae $ V $ band magnitude on metallicity, from various sources in the literature (the calibration of $ M_V(\text{HB}) $ by Harris (2010) has been adjusted, as suggested by Sandage, 1993, and Caloi et al., 1997). }
	\label{Fig:1}
\end{figure}

Figure 1 shows a number of calibrations of the $ M_V(\mathrm{RR}) - \mathrm{[Fe/H]} $ relationship from the literature. Statistical parallax (e.g. Gould \& Popowski 1998 and sources in Layden 1998) appears faint by other measures. The Baade-Wesselink method (Fernley et al. 1998) is also generally regarded as faint and Bono et al.'s (2007) theoretical calculation is also perhaps faint. Close agreement is found between the theoretical calibration of Catelan et al. (2004) and recent calibrations from sources based on a range of methodologies by Harris (2010) and Cacciari \& Clementini (2003) for $ \mathrm{[Fe/H]} <\sim -0.6 $, but, linear relations are known to fail for metallicities greater than $ \sim -0.5 $ dex. The `zero point' calibration using Hubble Space Telescope (HST) parallaxes for five RR Lyrae variables by Benedict et al. (2011) is significantly brighter. 

We finally adopted the calibration of Catelan et al. (2004) after establishing from dispersion that it is in better agreement than those of Harris (2010) and Cacciari \& Clementini (2003) with distances calculated from the range of methods used in our database. Thus, we adopted
\begin{equation}
M_V (RR) = 1.067 + 0.502 \mathrm{[M/H]} + 0.108\mathrm{[M/H]}^2.
\end{equation}
We used the conversion of Salaris et al. (1993),
\begin{equation} \label{eq:2.2}
\mathrm{[M/H]} = \mathrm{[Fe/H]} + \log (0.638\times 10^{[\alpha/\mathrm{Fe}]} + 0.362)
\end{equation}
assuming $ \mathrm{[\alpha/Fe]} = 0.3 $ (e.g., Carney 1996), together with the established relation of, e.g., Sweigart and Gross (1976) which was confirmed in theoretical models by Caloi et al. (1997) 
\begin{equation}
M_V (\mathrm{ZAHB}) = M_V(\mathrm{RR}) + \sim 0.06
\end{equation}and the empirical correction of Sandage (1993)
\begin{equation}
M_V (\mathrm{HB}) = M_V(\mathrm{ZAHB}) - 0.05(\mathrm{[Fe/H]} + 1.5) - 0.09 
\end{equation}
The error in equation (2.1) can be estimated from the calibrations of Harris (2010) and Cacciari and Clementini (2003), who found zero point errors $ \pm 0.049 $ mag and $ \pm 0.03 $ mag. The conversions given by equations (2.2), (2.3) and (2.4) are of the same order of magnitude as the error in the calibration. Errors in these corrections are absorbed into the total error in quadrature. 

We recalculated cluster distances when the source contained the value for one of these magnitudes using these relations together with updated metallicities from Saviane et al. (2012) when available. These are homogeneous with metallicities in H10, as the metallicity scale established by Carretta et al. (2009) is used. When the source contains an accurate determination of reddening we have used that figure. Otherwise, we used the value from H10. 

We calculated distances using the arithmetic mean of all distinct measurements in our database (the weighted mean is not justified as errors in the sources are not homogeneous and measurements may share systematic errors). The error stated in the online catalogue is given by
\begin{equation} \label{eq:2.5}
\epsilon=\frac{1}{n} \sqrt{\bar{\mathrm{\sigma }}^2 + \mathrm{Var}(R)(n-1)}
\end{equation}
where $ R $ is the heliocentric distance and $ \sigma $ is the error quoted in the source; $n$ is the number of measurements for a given cluster. This ensures that as the number of measurements increases the error estimate approaches the value calculated from the dispersion of the results. When no error was given in a source, we used a nominal error of 10\% (greater than most estimated errors) to calculate  in equation (2.5). With this estimate, three quarters of the population have errors below 4.3\%, and half have errors below 2.7\%. Nine clusters within 5 kpc of the Galactic Centre have distance errors of at least 10\%. These are clusters with substantial reddening on which few studies have yet be carried out. Removing these clusters makes little difference to the plots in equation (3) and does not affect our estimate of $R_0$ kpc. We have left them in the analysis reported here.  

\section{Halo Centroid}\label{Halo Centroid}
\begin{figure}
	\centering
		\includegraphics[width=0.47\textwidth]{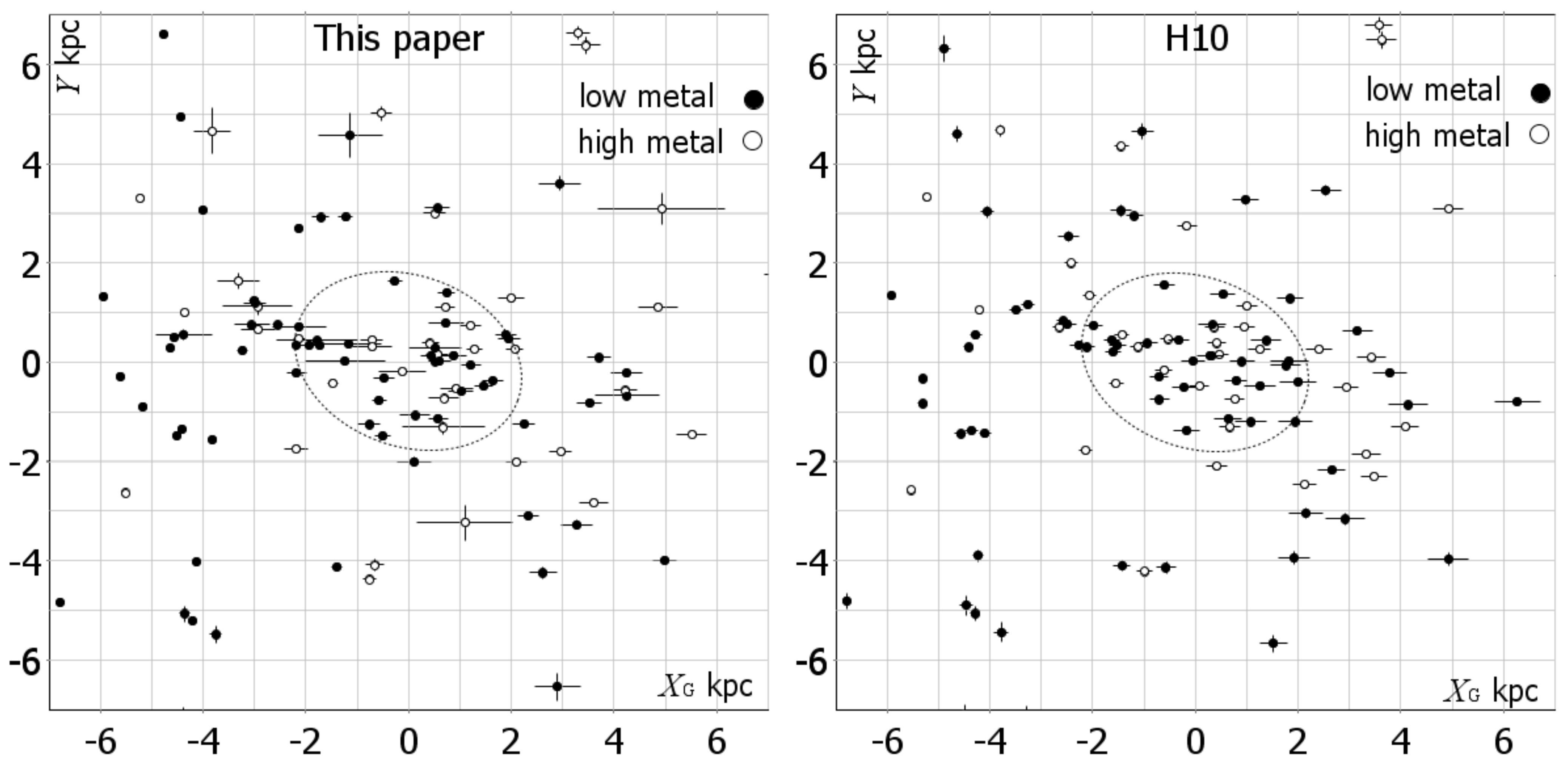}
	\caption{The $ XY $-distribution of globular clusters using distances in our database (left) and distances from H10 (right). Clusters with $ \text{[Fe/H]} \le 0.9 $ are shown with black circles and clusters with $ \text{[Fe/H]} > 0.9 $ are shown with open circles. The position of the bar/bulge is seen in the dense central region (roughly delineated by the dotted oval). Error bars in our database incorporate dispersion in measurements. Error bars for H10 are from equation 3 in Harris (2010) and do not allow for errors due to reddening. }
	\label{Fig:2}
\end{figure}
We plotted the distributions of clusters in the $ XY $ plane (figure 2) using distances in our database (left) and distances in H10 (right) for comparison. $X$ is toward the Galactic Centre, $Y$ is in the direction of rotation. A shift, $ X_G = X - 7.4 $ has been applied to show the Galactic Centre at the origin, as estimated from the halo centroid. Clusters with $ \mathrm{[Fe/H]} \le -0.9 $ are shown with black circles and clusters with $ \mathrm{[Fe/H]} > -0.9 $ are shown with open circles. The position of the bulge/bar is seen in both databases as an overdense region with major axis at an angle about 20\textdegree{} to our line of sight (this is not a good estimate of the bar angle because of the small sample size). There is little evidence of regions of extinction. It might be expected that a small number of globular clusters close to the Galactic plane are hidden by dust, but recent all-sky surveys such as 2MASS and SDSS have not uncovered many new clusters, and, given the ready availability of digital search algorithms, and that the position and approximate angle of the bar are clearly seen in figure 2 it seems improbable that there might exist large numbers of yet undiscovered clusters which would substantially change our estimate of $R_0$ kpc. If extinction in the central region were to hide enough undiscovered clusters to substantially alter the visible position of the bar in figure 2, then a clear asymmetry would be expected in the distribution away from the Galactic Centre, but this is not seen.
\begin{figure}
	\centering
		\includegraphics[width=0.47\textwidth]{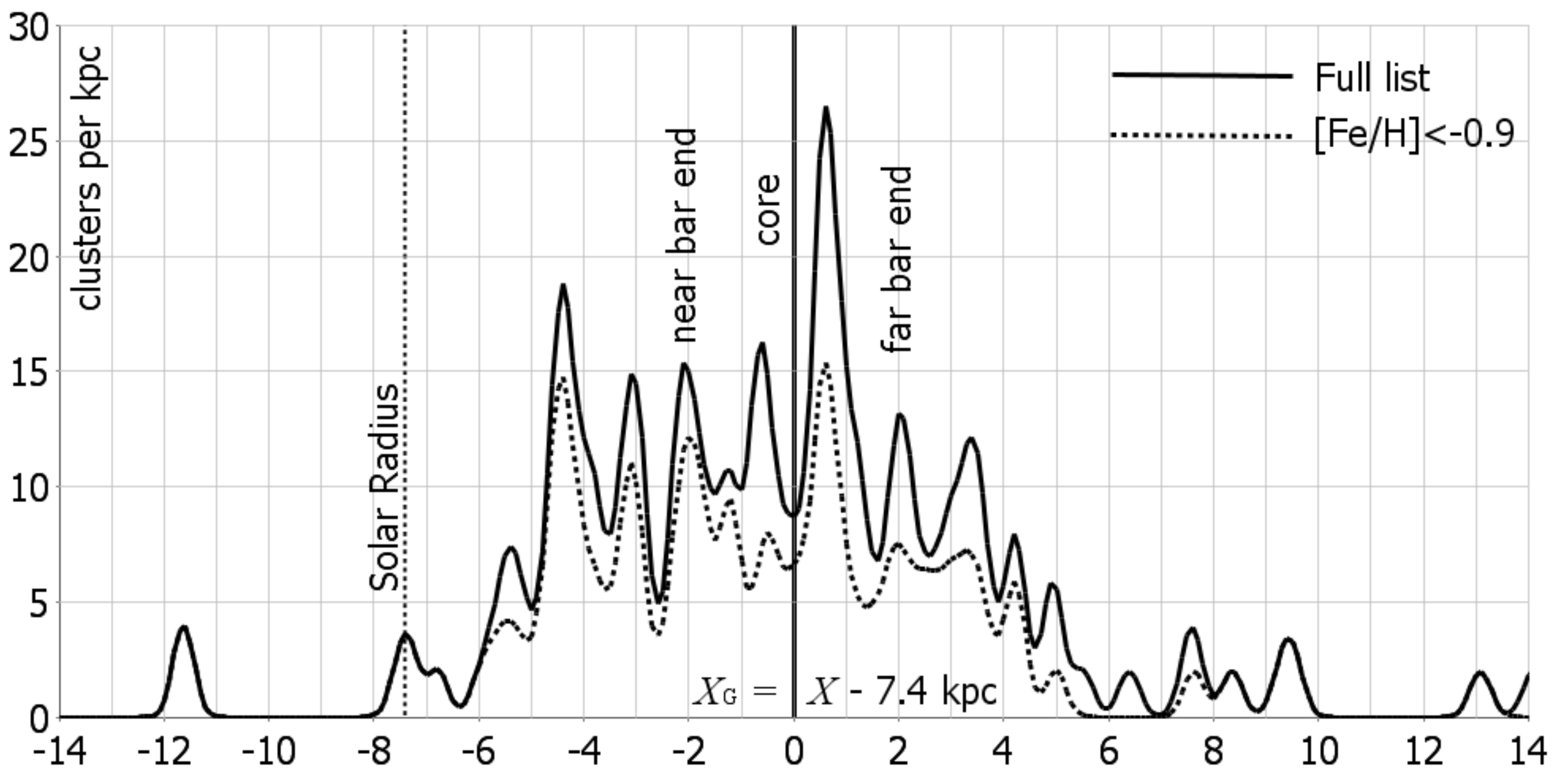}
	\caption{Distribution of cluster positions on the $ X $-axis, from the Sun to the Galactic Centre, for the full database, and for clusters with metallicities $ \text{[Fe/H]} \le 0.9 $ (dotted). The scale has been shifted by 7.4 kpc, to show the Galactic Centre at the origin. }
	\label{Fig:3}
\end{figure}\begin{figure}
	\centering
		\includegraphics[width=0.47\textwidth]{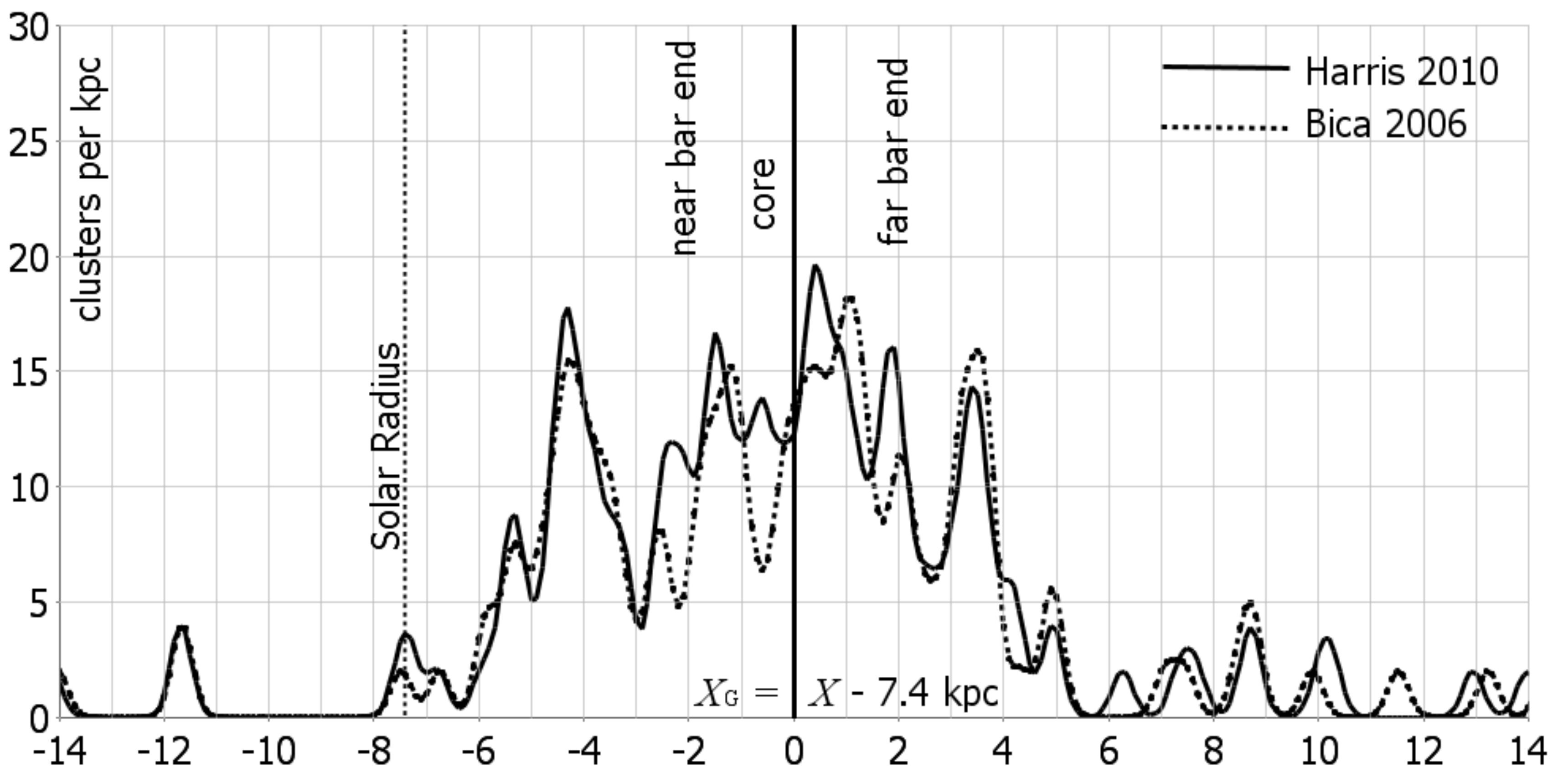}
	\caption{As figure 3, but using distances from H10 and from B06. }
	\label{Fig:4}
\end{figure}

We plotted the distribution of globular clusters on the $X$-axis by replacing each discrete data point with a Gaussian with standard deviation 0.2 kpc centred at the datum, and forming the sum (figure 3). This method, called Gaussian smoothing or kernel estimation (Silverman 1986), is an alternative to binning for finding a distribution function, but gives a more accurate result with better resolution, by finding a form of ``moving sum'' (c.f. the more familiar concept of moving average). The chosen standard deviation is a smoothing parameter for the distribution function. Also called bandwidth, the smoothing parameter loosely corresponds to bin size, and suppresses fluctuations over distances less than about twice the standard deviation. We chose the smoothing parameter, 0.2 kpc, by applying Silverman's rule of thumb after approximating the properties of individual major peaks with normal distributions (Silverman's rule of thumb states that if the underlying distribution is normal with standard deviation $ \tilde{\sigma} $ and the population size is $n$, then the optimal smoothing parameter is $ 1.06\tilde{\sigma}n^{-0.2} $). In practice the method is not critically dependent on the choice of an optimal smoothing parameter. One arrives at a similar value from a subjective judgement of the balance between removing random fluctuations while retaining structural information.

The plot shows a remarkable amount of structure. The structured form of the distribution, together with the fact that a number of globulars may still be undiscovered, especially on the far side of the bulge, means that traditional statistical measures like mean, mode and median are not robust indicators of the centroid. For each of the distributions seen in figure 3 and figure 4, it is clear that the centre of the distribution is a little above 7 kpc from the Sun. The central trough is greater than the median (7.0 kpc) but a reduction in frequency on the far side of the bulge is likely, due to undiscovered clusters. Undiscovered clusters would not be expected to substantially alter the position of troughs and peaks in the distribution because the statistical properties of a sample can usually be expected to reflect the properties of a population. 

The same underlying structure is seen in the low metallicity population. The Galactic Centre is seen as a trough at a distance of 7.4 kpc, close to the centre of the distribution. The scale on the $X$-axis has been shifted to show the trough at the origin. We restricted the population to 134 clusters within 20 kpc of the Galactic Centre, based on $ R_0 = 7.4 $ kpc. This made little difference to the central part of the diagram. A further restriction to 10 kpc from the Galactic Centre also made very little difference to the central part of the diagram.

For comparison we plotted the distribution with distances given by H10 and by B06 (figure 4). Similar features are seen. The central trough lies at near the same position using distances from H10, and a little nearer using B06. Since random effects are more likely to obscure regular structure than to create it, we believe the clearer structure in figure 3 results from the improved accuracy of the newer database. 

The central trough may be understood because a cluster near the Galactic Centre would interact strongly with the central density cusp and would not be expected to survive. Clusters on larger orbits are limited by angular momentum as to how close they can come to the Galactic Centre and most time is spent near to apocentre. Because of this, and because the angle of the bar means that clusters will usually not be aligned with the Galactic Centre on the line of sight from the Sun, a trough is seen in the distribution. 

The position of peaks either side of the central trough is highly symmetric, and corresponds to the known position of a bar of length a little over 4 kpc, in agreement with other estimates (Francis \& Anderson 2012a, Nataf 2013, Babusiaux and Gilmore 2005, Vanhollebeke et al. 2009 and references therein). The peaks to either side of the central trough can be interpreted as tangencies to highly eccentric orbits of globular clusters within the bar, together with an overdensity of clusters near apocentre where clusters are moving slowly and spend most time. The visibility of the symmetrical structure of the bar supports the identification of the central trough with the Galactic Centre.

The position of the central trough, together with the symmetry of the peaks corresponding to the bar, gives a distance, $ R_0 = 7.4 \pm 0.2 $ kpc, to which can be added whatever systematic is contained in estimates of cluster distances, due to uncertainties in absolute magnitude and in estimates of the effect of reddening. The errors in cluster distances in our database are below 4\% for 75\% of clusters, and most cluster distances are not greatly affected by reddening.

The troughs to either side of the central trough are not realistic candidates as markers for the Galactic Centre because they are outside the range of estimates published since 2000 (section 6), and because they are substantially removed from the centre of the distribution; the choice of either of these would mean assuming an unrealistic number of as yet undiscovered clusters.

Random errors will largely cancel out from the statistical properties of the distribution, so the main source of error is likely to be systematic, and dependent on the calibration of horizontal branch magnitudes which are use to determine most (but not all) cluster distances. To estimate systematic errors, we considered the calibrations of Harris (2010) and Cacciari and Clementini (2003), for which the zero point errors are respectively $ \pm 0.049 $ mag and $ \pm 0.03 $ mag. At the distance of the Galactic Centre these figures give a systematic error of 0.17 kpc and 0.10 kpc. We have therefore estimated a net systematic error of $ \pm 0.2 $ kpc.

\section{Calibration of the red clump}\label{Calibration of the red clump}
To calibrate the magnitude of red clump stars in Hipparcos with HNR parallaxes we removed stars flagged as variable or multiple, and components of multiple stars from the Catalog of Components of Double \& Multiple Stars (Dommanget \& Nys 2002) and The Washington Visual Double Star Catalog (Mason et al. 2001-2010) (for which magnitudes and parallaxes are less accurate and may be subject to systematic errors). Initially, we restricted to stars with parallax errors better than 25\%, and Hipparcos goodness of fit flag |F2| < 5. For the purpose of calibration of the $K$ band only, we restricted to stars with 2MASS quality flags A-D in each of the $J$, $H$, and $K$ bands (a valid magnitude has been obtained). We applied bias corrections described in Francis (2013) due to non-linearity of the distance modulus with respect to parallax, and non-uniformity of the stellar distribution perpendicular to the Galactic plane. 

We dereddened the Hipparcos population outside 100 pc (the local bubble) for Galactic latitudes greater than $ b = 9.7 $\textdegree{} using the maps of Burstein \& Heiles (1978, 1982), together with Bahcall \& Soneira's (1980) formula,
\begin{equation}\label{eq:4.1}
A_d(b)=A_{\infty}(b)(1-\exp(\frac{-|d\sin b|}{h})),
\end{equation}
where $ A_{\infty}(b) $ and $ A_d(b) $ are total absorption at infinity and at stellar distance, $ d $; $ A_{\infty}(b) =3.1E_{\infty}(B-V)$ is found from the reddening map; $ h = 125 $ pc is the adopted scale height for interstellar dust (Marshall et al. 2006). We excluded stars outside 100pc with latitudes less than 9.7\textdegree{} which cannot reliably be dereddened by this method. Absorption in each magnitude is found using, $ A_B = 4.325E(B - V) $, $ A_V = A_{Hp} = 3.1E(B - V) $, $ A_I = 1.962E(B - V) $, $ A_J = 0.902E(B - V) $, $ A_H = 0.576E(B - V) $, $ A_K = 0.367E(B - V) $ (Schlegel, Finkbeiner \& Davis 1998). Burstein \& Heiles map is of lower resolution than that of Schlegel et al., but is compatible for Galactic latitudes above $ b = 9.7 $\textdegree{} and is empirically based on the reddening of other galaxies, rather than calculated theoretically from dust maps. This choice is likely to have little practical impact.

Parallax errors in HNR are less for bright stars, so a boundary on parallax error will generate a bias toward bright stars. It is therefore necessary to impose a strict limit on distance, rather than a bound on parallax error. We compared results after restricting the population to distances less than 150 pc (max parallax error 19.6\%; mean 5.5\%) and 100pc (max parallax error 11.9\%, mean 2.9\%), finding no appreciable difference in statistical properties. We elected to use statistics from the larger population because statistical errors are smaller.

Paczy\'{n}ski and Stanek (1998) required that the value of Hipparcos flag H42 is taken from \{A,C,E,F,G\}, meaning that the star has one or more direct measurements of the $I$-band. The Hipparcos catalogue gives inferred values of $ V - I $ for all stars, but one would expect these to be less accurate than values for which $I$ has been measured. Following Udalski (2000), calibrations of $ M_K(\mathrm{RC}) $ by Alves (2000) and Groenewegen (2008) also allowed H42 = ``H''. Selecting on the H42 flag may introduce a selection bias for bright stars, since these are more likely to have been chosen for measurements of the $I$ band. To ascertain the presence of a selection bias we calibrated $ M_I(\mathrm{RC}) $ and $ M_K(\mathrm{RC}) $ both with and without the restriction on H42, and we also calibrated $ M_{Hp}(\mathrm{RC}) $ as a control. 

The calibration of $ M_K(\mathrm{RC}) $ presents a separate problem, loss of accuracy and possible systematic error caused by saturation of the detectors for near stars for which accurate parallaxes are available (Cutri et al. 2003, Skrutskie et al. 2006). Only a small number of giants within 150 pc have quality index q\underline{ }JHK = ``AAA''. Alves (2000) used $K$-magnitudes from the Two Micron Sky Survey (TMSS, Neugebauer \& Leighton 1969) finding agreement with other $K$-band measurements, to within $ \sim0.01-0.02 $ mag. Groenewegen (2008) used 2MASS measurements when the quality flag, q\underline{ }K = ``A'', and otherwise used non-saturated DENIS or TMSS magnitudes after transforming to the 2MASS system (subtracting 0.011 from DENIS and $ \sim 0.02 $ from TMSS). Alves (2000) obtained $ M_K(\mathrm{RC}) = -1.61 \pm 0.03 $ mag on the TMSS system, but Groenewegen (2008) found $ M_K(\mathrm{RC}) = -1.54 \pm 0.04 $ on the 2MASS system), almost 0.1 mag less bright, which he attributed to selection bias. A recent recalibration using new measurements of $K$ mag by Laney et al. (2012) found $ M_K(\mathrm{RC}) = -1.61 \pm 0.02 $ mag, in agreement with Alves and based on a subset of the same population in which selection bias has been identified. 

Saturation in 2MASS does not necessarily invalidate measurements because the project was able to measure the rate of photon detection prior to saturation for many sources. Although errors are greater for quality indices above ``A'', usable magnitudes are obtained for quality indices up to ``DDD'', meaning that valid magnitudes have been obtained. Allowing quality indices to ``EEE'' included almost all red clump giants within 150 pc but led to a less bright peak magnitude, by 0.04 mag. This could be thought to mean that this population contains a significant number of stars with systematically high magnitudes resulting from saturation of the detectors. We restricted to quality indices ``DDD'' and better. We compared statistics with other cuts (25\% plx errors, 10\% plx errors, distance less than 100pc) and did not find a significant difference in the peak magnitudes. We could not find any reason in the data to think that allowing quality indices to ``DDD'' introduces a significant systematic bias. 
\begin{figure}
	\centering
		\includegraphics[width=0.47\textwidth]{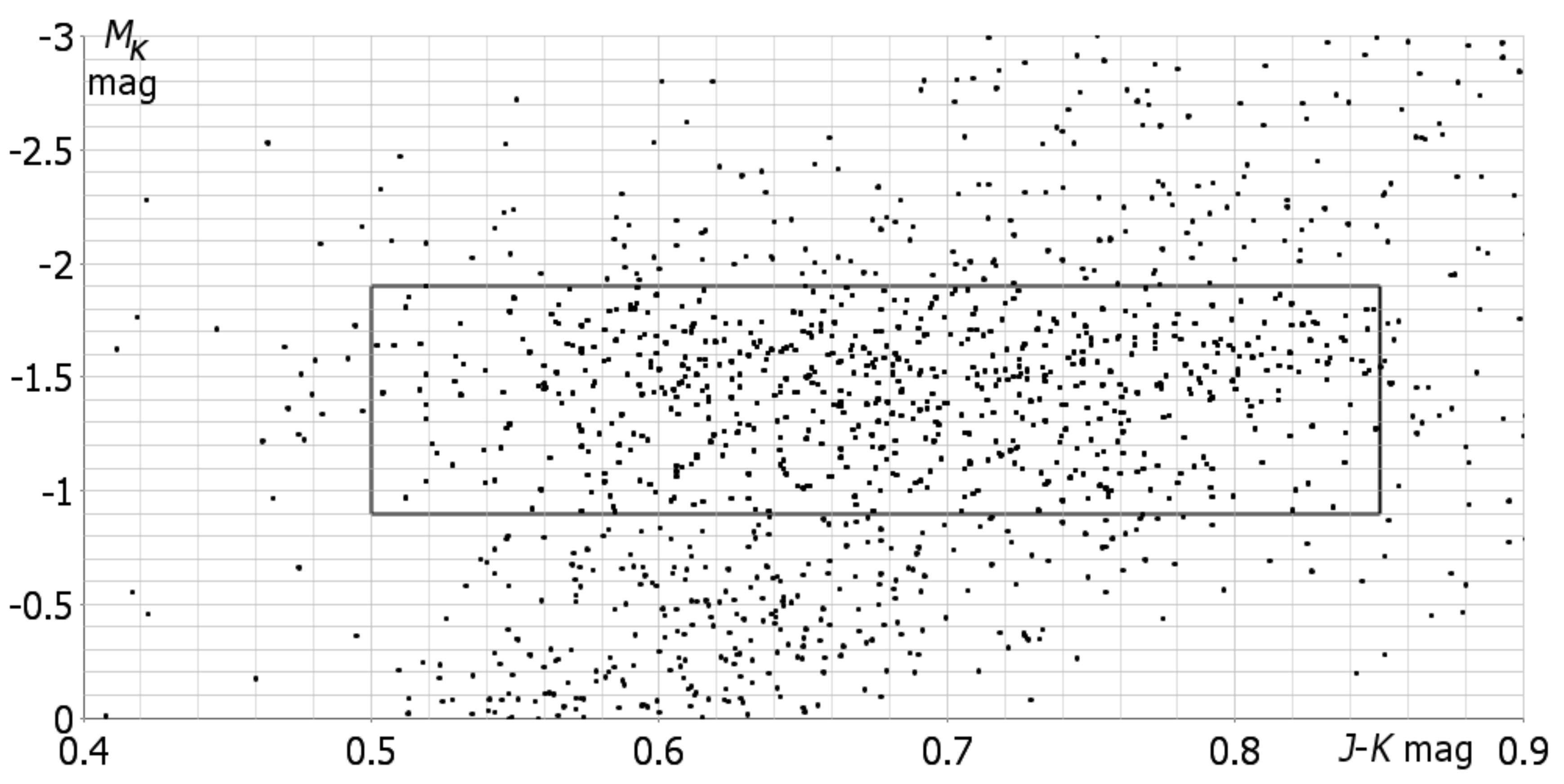}
		\includegraphics[width=0.47\textwidth]{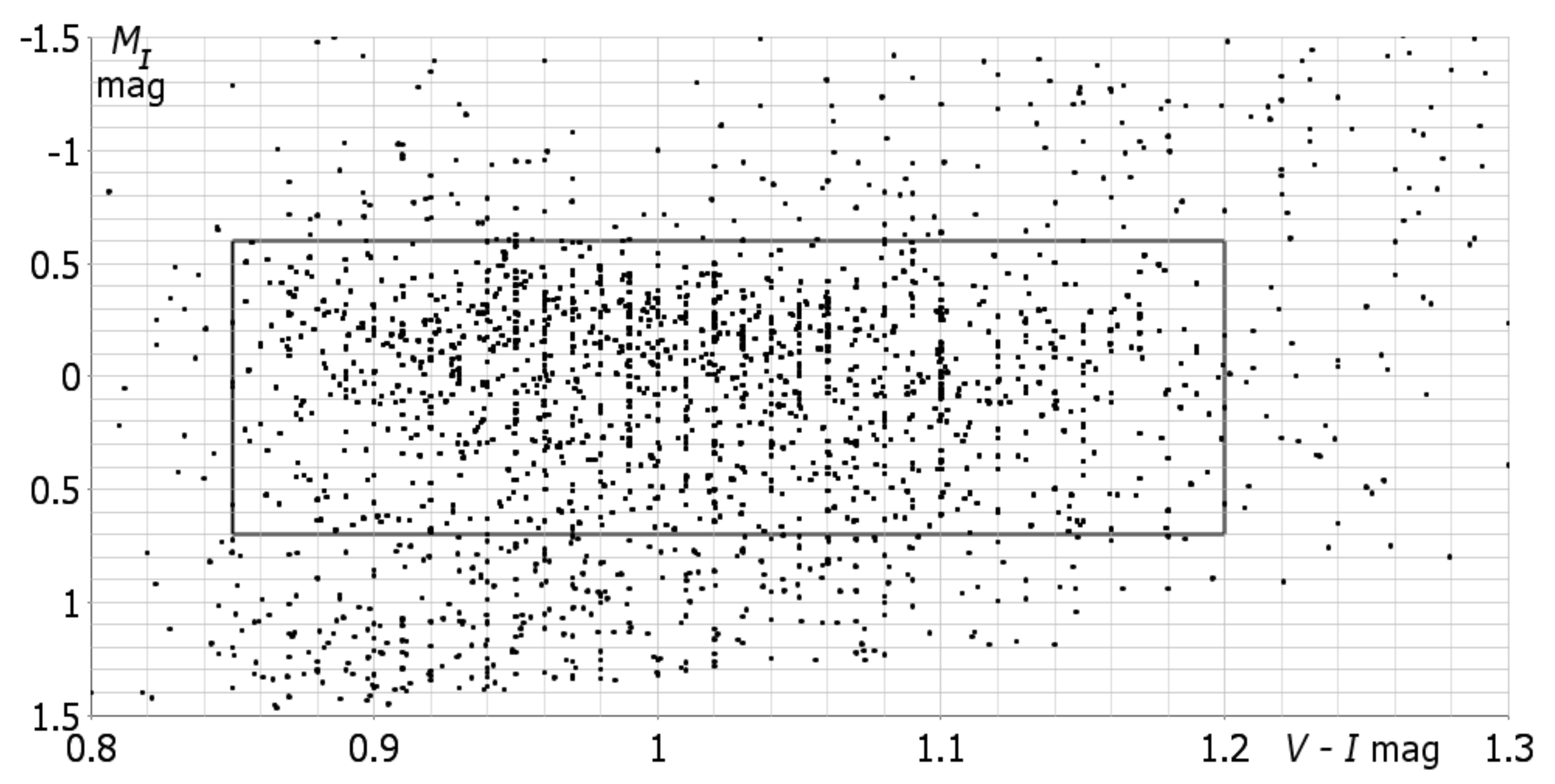}
		\includegraphics[width=0.47\textwidth]{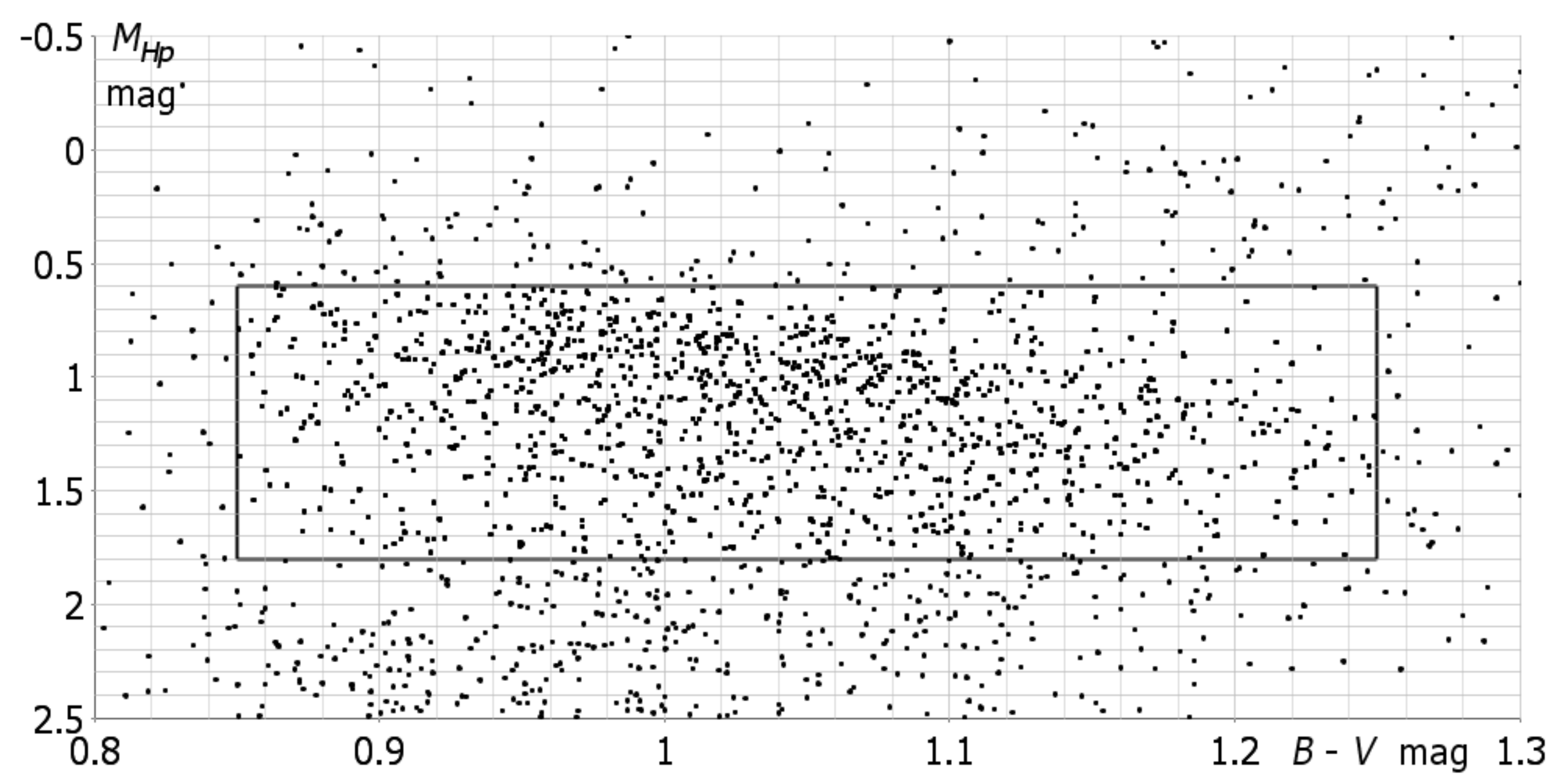}
	\caption{The red clump as defined for $ K $, $ I $ and $ Hp $ bands for stars within 150 pc. For the $ K $ band, stars have 2MASS quality index ``DDD'' or better. }
	\label{Fig:5}
\end{figure}

We plotted colour magnitude diagrams for $ M_{Hp} $ against $ B - V $, $ M_I $ against $ V - I $, and $ M_K $ against $ J - K $ (figure 5). For the $I$ band calibration we restricted the population to the red clump using colours in the range $ 0.85 < V - I \le 1.2 $ and absolute magnitudes in the range $ -0.6 < M_I \le 0.7 $. For the $ K $ band calibration we restricted the population to the red clump using colours in the range $ 0.5 < J - K \le 0.85 $ and absolute magnitudes in the range $ -1.9 < M_K \le -0.9 $. For the control ($Hp$) population we restricted to the red clump using colours in the range $ 0.85 < B - V \le 1.25 $ and absolute magnitudes in the range $ 0.6 < M_{Hp} \le 1.8 $.
We found normalised distributions in each band by replacing each discrete data point with a Gaussian with standard deviation 0.05 centred at the datum, forming the sum and dividing by the number of data points (figure 6). This is more precise than binning for calculating the peak of the distribution. The smoothing parameter, 0.05 mag, is less than suggested by Silverman's rule of thumb (0.07-0.10 depending on the plot) but is justified because the distributions are more sharply peaked and broader at the base than normal distributions, and because 0.05 mag is sufficient to damp out random fluctuations.

The $Hp$ and $I$ bands are brighter by a similar amount with the restriction on H42. We concluded that the H42 flag introduces a selection bias, and that there is no significant systematic error from ignoring the H42 flag. Since a larger sample with greater random errors is preferable to a smaller sample with a systematic error, we elected to use the recalibration with no restriction on H42 in corrected estimates of $R_0$ kpc.

\begin{figure}
	\centering
		\includegraphics[width=0.47\textwidth]{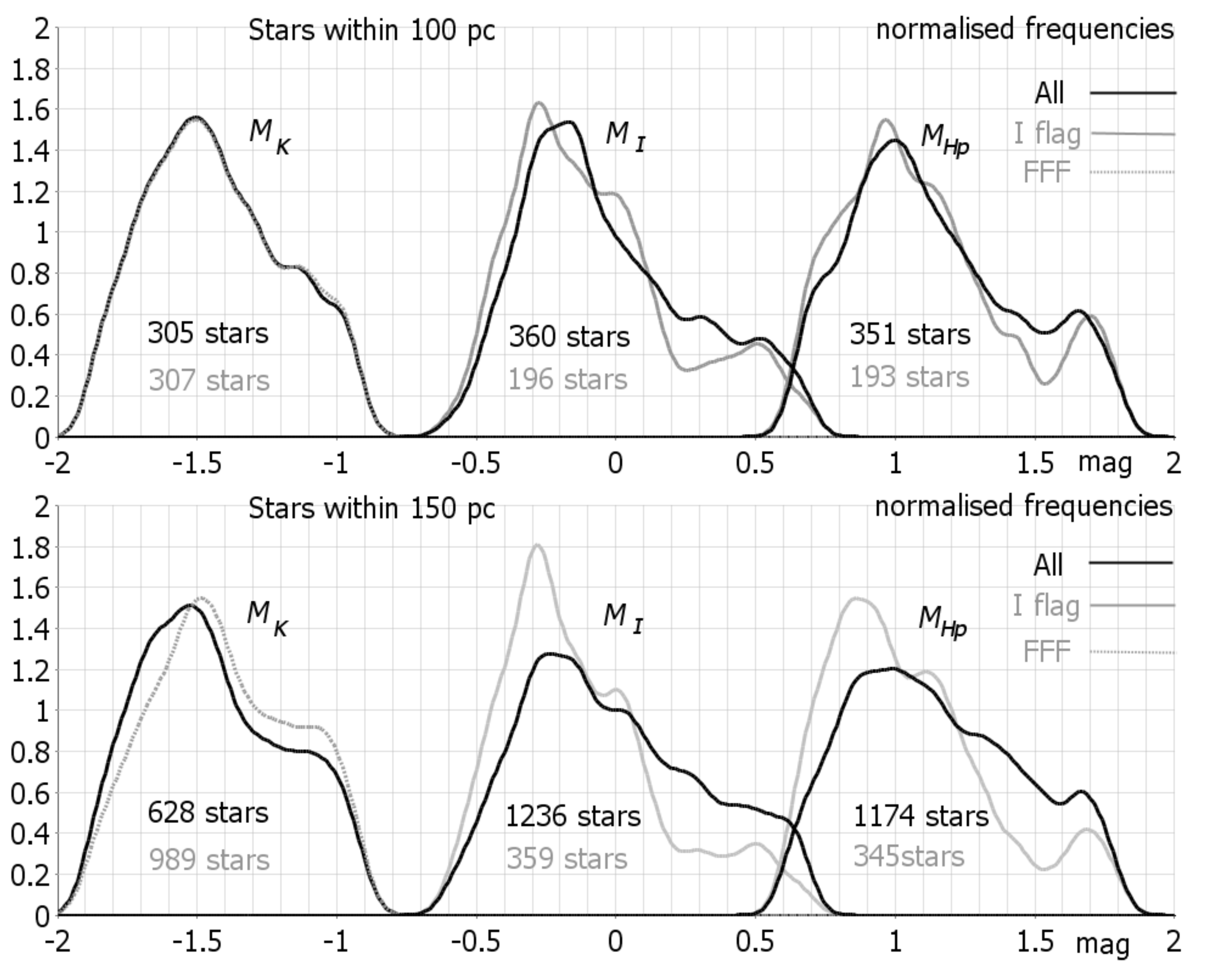}
	\caption{Normalised magnitude distribution of the red clump for $ K $, $ I $ and $ Hp $ bands for stars within 100 pc and 150 pc. For the $ K $ band, stars with 2MASS quality index ``DDD'' or better are shown in black. The dashed grey line shows stars with quality index ``EEE'' or better. For the $ I $ and $ Hp $ bands, stars with H42 in \{A,C,E,F,G\} are shown in grey, and the full distribution is shown in black. }
	\label{Fig:6}
\end{figure}
\begin{table}\label{(table1)}
\begin{small}
\begin{flushleft}
\begin{tabular}{lrrrrr} 
\textbf{band} &	\textbf{distance} &	\textbf{count} & \textbf{$ M $ peak} & \textbf{$ M $ mean} &	\textbf{error}\\
$ K  $ & $ < 150 pc $ & $ 628		$ & $ -1.53	$ & $-1.424	 $ & $ 0.010$  \\
$ K  $ & $ < 100 pc $ & $ 305		$ & $ -1.51	$ & $-1.393	 $ & $ 0.014$  \\
$ I  $ & $ < 150 pc $ & $ 1 236	$ & $ -0.24	$ & $0.043	 $ & $ 0.009$  \\
$ I  $ & $ < 100 pc $ & $ 360		$ & $ -0.17	$ & $0.032	 $ & $ 0.016$  \\
$ Hp $ & $ < 150 pc $ & $ 1 174	$ & $ 0.99	$ & $1.180	 $ & $ 0.009$  \\
$ Hp $ & $ < 100 pc $ & $ 351		$ & $ 1.00	$ & $1.201	 $ & $ 0.016$  \\
\end{tabular}
\caption{Peak and mean magnitudes of the red clump in the $ K $, $ I $ and $ Hp $ bands. No constraint on H42 is applied. $ K $ magnitudes use stars with 2MASS quality index ``DDD'' or better. Mean magnitudes are corrected for the Trumpler-Weaver bias.}
\end{flushleft}
\end{small}
\end{table}

Trumpler and Weaver (1953) commented on a selection bias affecting the mean parallax distance of a population within a sphere of given radius. The number of stars with true distances greater than this radius which appear in the sample due to parallax error will exceed the number with true distances inside the sphere whose parallax errors remove them from the sample, because the volume of the error shell outside the sphere is greater than the volume of the error shell inside the sphere. The consequence is that the true mean distance of the sample is greater than mean parallax distance. When a distance limited sample has been used for calibration, the Trumpler-Weaver bias will generate a systematic error in luminosity distances. Francis (2013) calculated the Trumpler-Weaver bias,
\begin{equation}\label{eq:4.2}
\Delta M = -5.8 (\overline{\sigma / \mathrm\pi})^2.
\end{equation}
Although equation (4.2) is calculated with a uniform stellar distribution, it does not depend on population density, and is therefore independent of direction (the bias applies to stars within a solid angle). The bias will be affected by a radial density gradient. 

Each of the distributions in figure 6 shows a positive skew, such that the peak magnitude is brighter than the mean. The distribution function is determined by the age and metallicity distribution of the population (Girardi and Salaris 2001). Consequently it is important to make an accurate determination of peak magnitude. 

The Trumpler-Weaver bias also contributes to the skewness of the distributions seen in figure 6. Stars removed from the sphere by parallax error have the same magnitude distribution as the population, whereas stars which are brought into the sphere are really further away, and appear less bright than they actually are. This boosts the number of stars in the right hand tails of the distributions in figure 6. Thus the Trumpler-Weaver bias shifts the mean, but it does not shift the position of the peak. We calculated the value of the Trumpler-Weaver bias for each population (using average parallax error according to equation (4.2)), finding $ \Delta M = -0.031 $ mag for a 150 pc radius and $ \Delta M = -0.008 $ for a 100 pc radius.

We used simple linear regression to determine mean magnitudes in each band as a function of colour. We found from 1 236 stars within 150 pc that $ M_I $ has a dependency on colour with slope $ -3.5 $, significant at 99.8\% (from Student's $ t $-test). We found no significant slope in $ M_K $ against colour from 628 stars. $ M_I $ has no measurable dependency on metallicity. $M_K$ has a metallicity slope $ -0.1 $ mag/dex, significant at 87\%, in agreement with the population correction predicted by Salaris and Girardi (2002).

As described by Francis (2013, section 2.5, magnitude bias) we applied a small correction to the mean magnitude ($ \sim 0.05 $ magnitude) in order to minimise the sum of squared differences between luminosity distances and parallax distances for the Hipparcos population. Magnitude bias arises in the calculation of the expected distances of stars from luminosity distances, and depends upon the real luminosity distribution and on the non-linear relationship between distance and magnitude. It is removed by minimising the sum of squared differences between luminosity and parallax distances for the calibration sample. It does not directly affect the calibration of peak magnitudes used in our determination of $R_0$, but the difference between peak and mean magnitudes is indicative of a systematic error which can arise in distance determinations using the red clump. Results are shown in table 1, after correcting mean magnitudes for the Trumpler-Weaver bias using equation (4.2).

The usual method of finding the peak in the magnitude distribution is to fit to fit a polynomial background plus a Gaussian model to the binned distribution function (e.g. Groenewegen 2008). The role of the background distribution is to remove the skew wings of the distribution function, such that the Gaussian is fitted to the peak. In consequence, our result is not directly comparable, but we believe the method followed here, finding the maximum of the smoothed distribution function, gives a more accurate and precise estimate. The peak $K$-band and $I$-band magnitudes for stars within 150 pc, $ M_K(\mathrm{RC}) = -1.53 \pm 0.01 $ and $ M_I(\mathrm{RC}) = -0.24 \pm 0.01 $ mag, are within 0.02 mag of the values found by Groenewegen, $ M_K(\mathrm{RC}) = -1.54 \pm 0.04 $ and $ M_I(\mathrm{RC}) = -0.22 \pm 0.03 $ mag but our analysis has led to smaller errors. This difference in magnitudes corresponds to a difference of less than 0.1 pc in estimates of the distance to the Galactic Centre, which is unimportant. Of greater importance is the difference between the peak magnitudes and the mean magnitudes seen in table 1, because this has a direct bearing on the treatment given to the distribution of stars in the bulge (section 6).

\section{An estimate of $R_0$ from 2MASS}\label{An estimate of $R_0$ from 2MASS}
Based on the calibration of the peak magnitude, we calculated the distribution in eight sectors at latitudes $ b = \pm 9.7$-$9.8 $\textdegree, $ b = \pm 10.4$-$10.5 $\textdegree, $ b = \pm 10.9$-$11.0 $\textdegree{} and $ b = \pm 12.0$-$12.1 $\textdegree. These latitudes are chosen, together with the $ K $-band, to minimise errors due to reddening. We found a first estimate of luminosity distance for each of the sample stars using,
\begin{equation}\label{eq:5.2}
R = 10^{(K-M_K)/5+1}.
\end{equation}
We carried out reddening corrections using equation (3.1) iteratively. Thus, distances are first calculated using the value of absorption at infinity, absorption is calculated for these distances using equation (3.1), then distances are recalculated using the corrected magnitude. The procedure is repeated until convergence is achieved to the required accuracy (in practice one iteration is sufficient at these distances and latitudes). 
\begin{figure}
	\centering
		\includegraphics[width=0.235\textwidth]{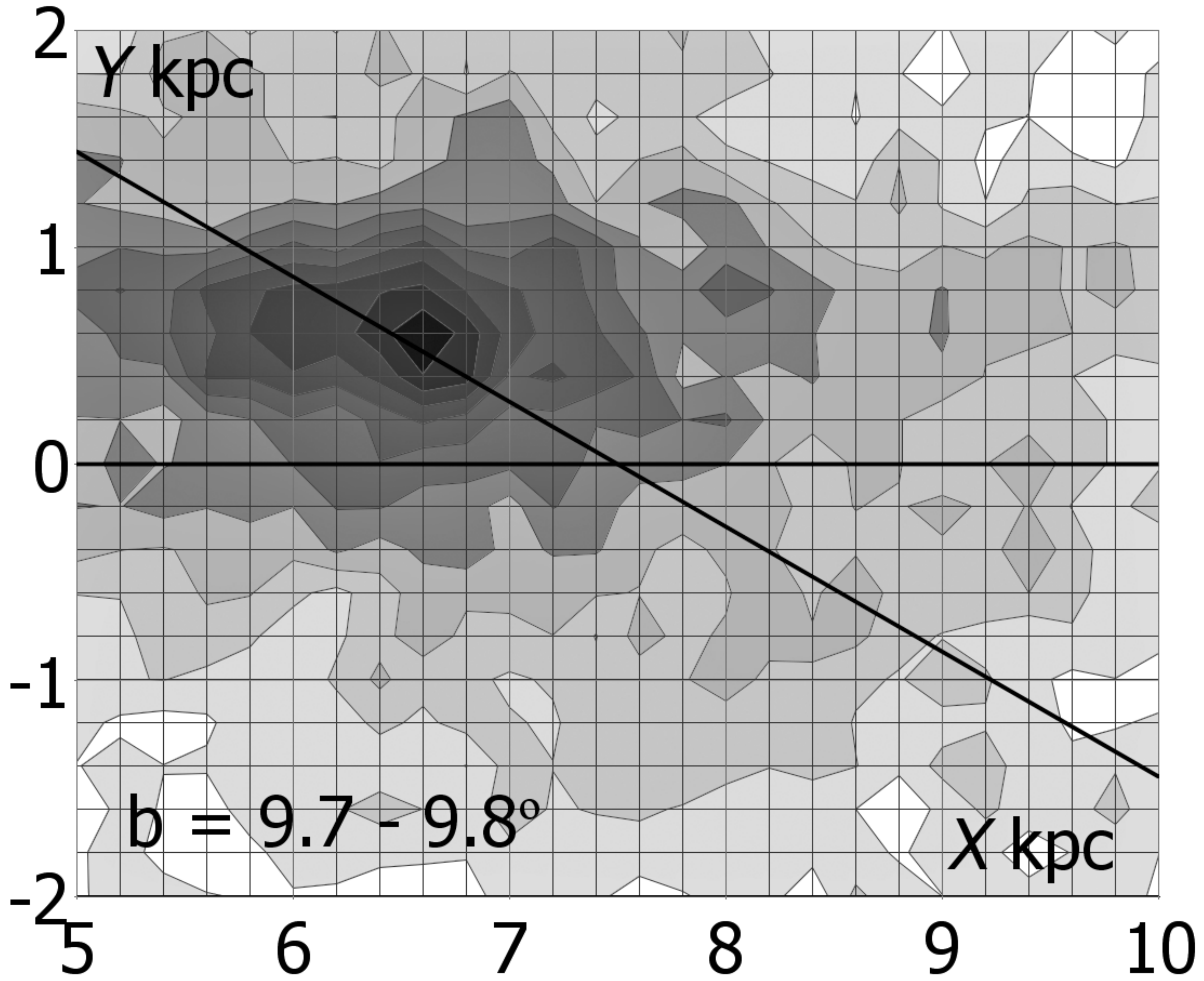}
		\includegraphics[width=0.235\textwidth]{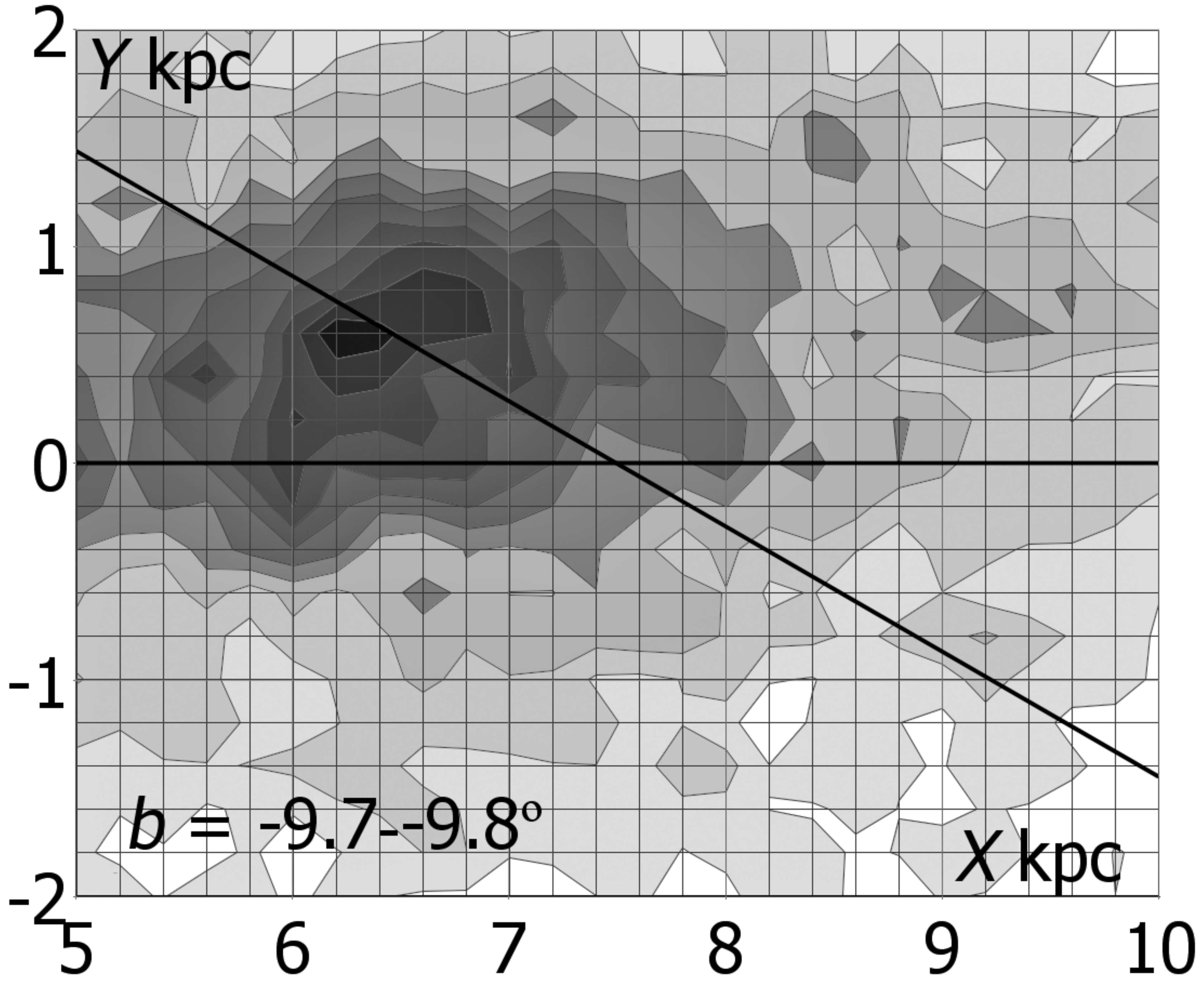}
		\includegraphics[width=0.235\textwidth]{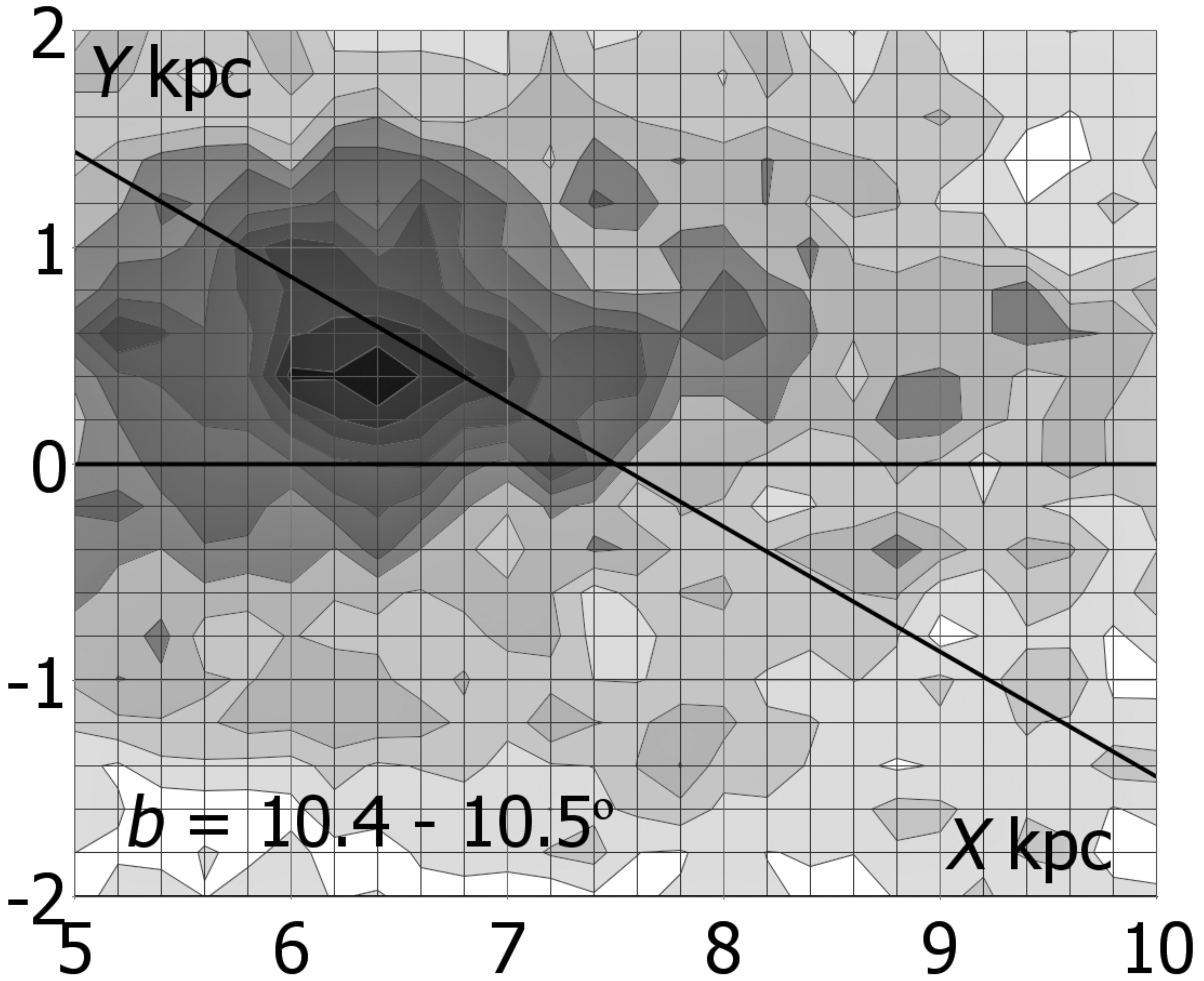}
		\includegraphics[width=0.235\textwidth]{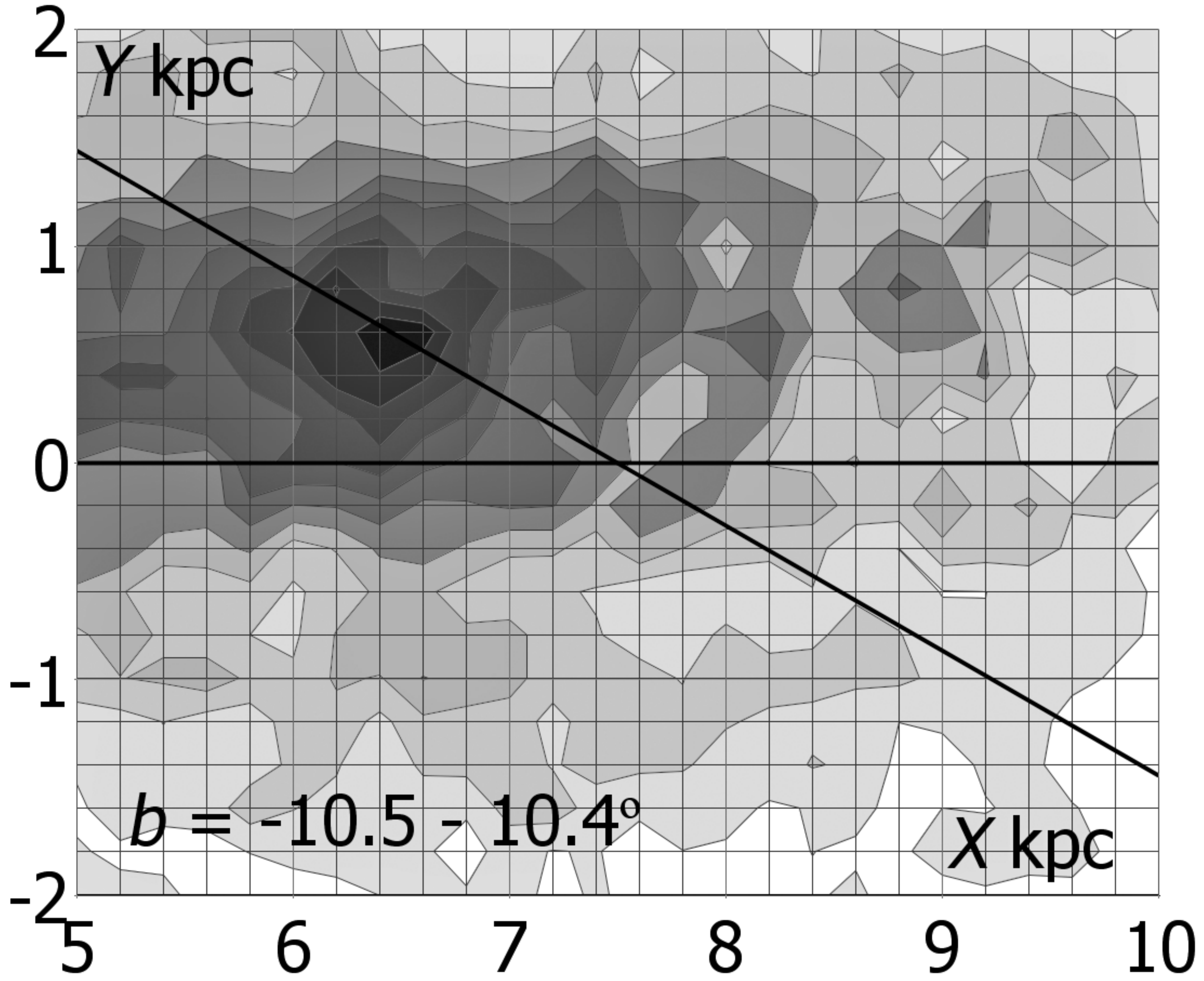}
		\includegraphics[width=0.235\textwidth]{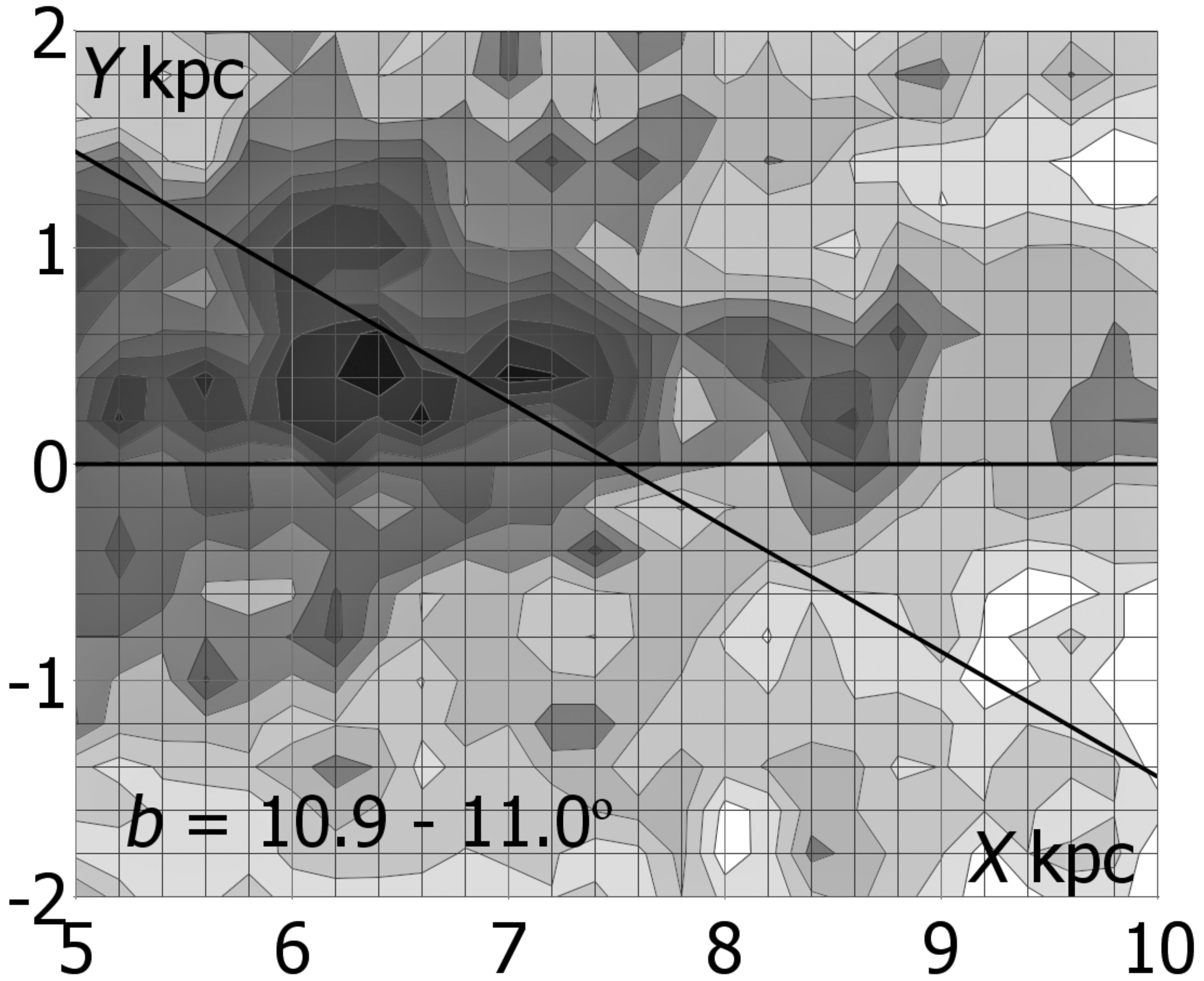}						\includegraphics[width=0.235\textwidth]{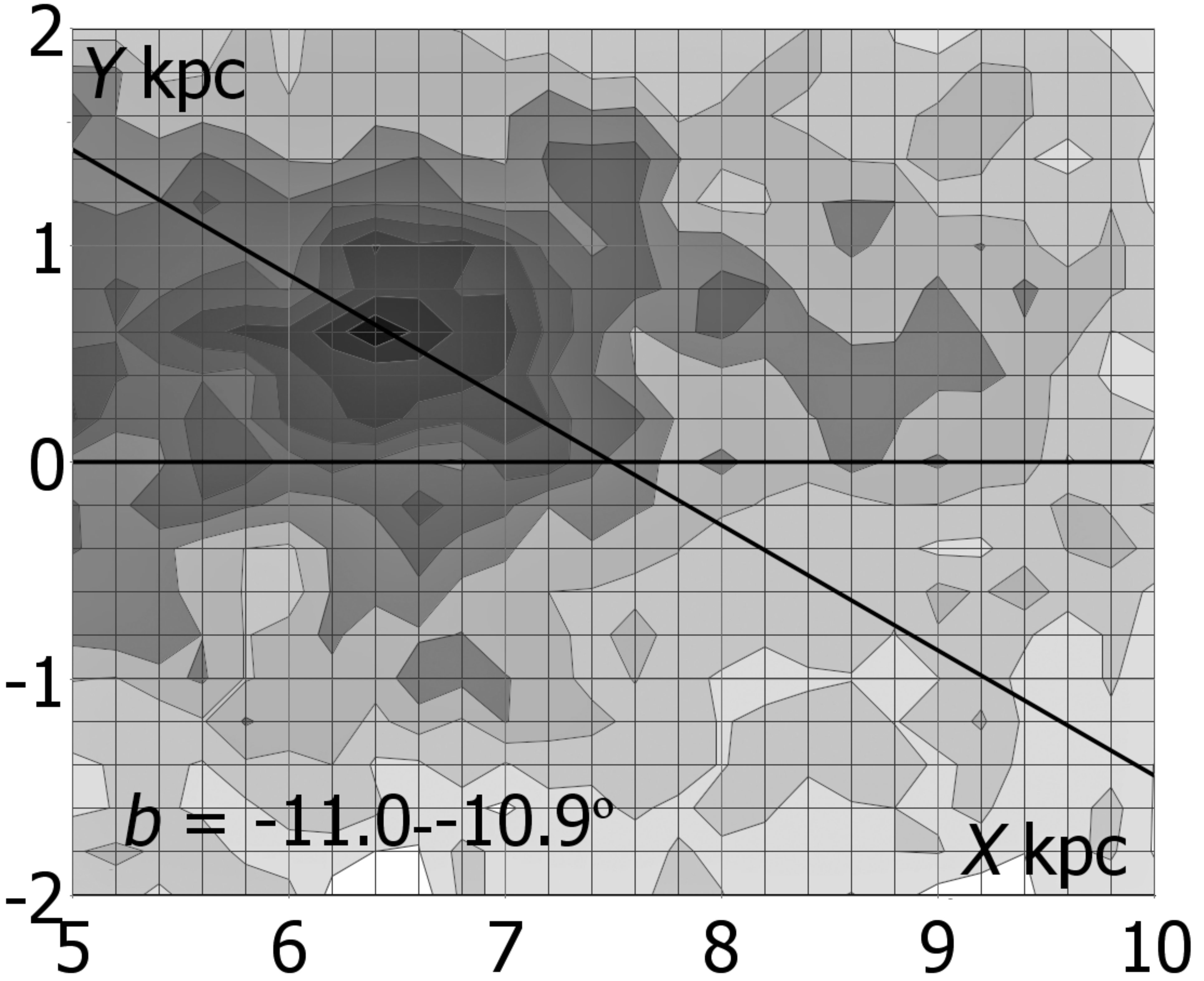}				
		\includegraphics[width=0.235\textwidth]{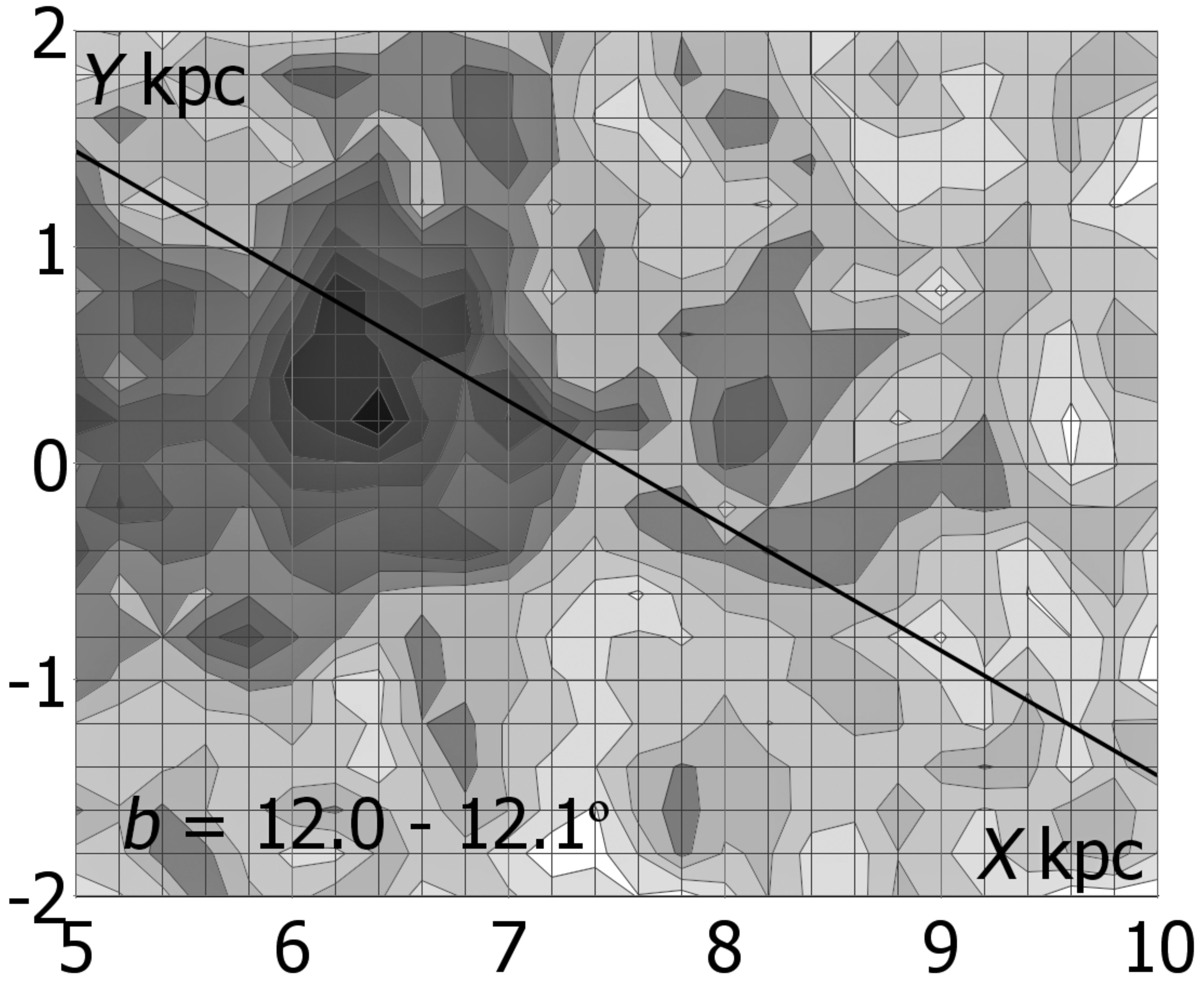}
		\includegraphics[width=0.235\textwidth]{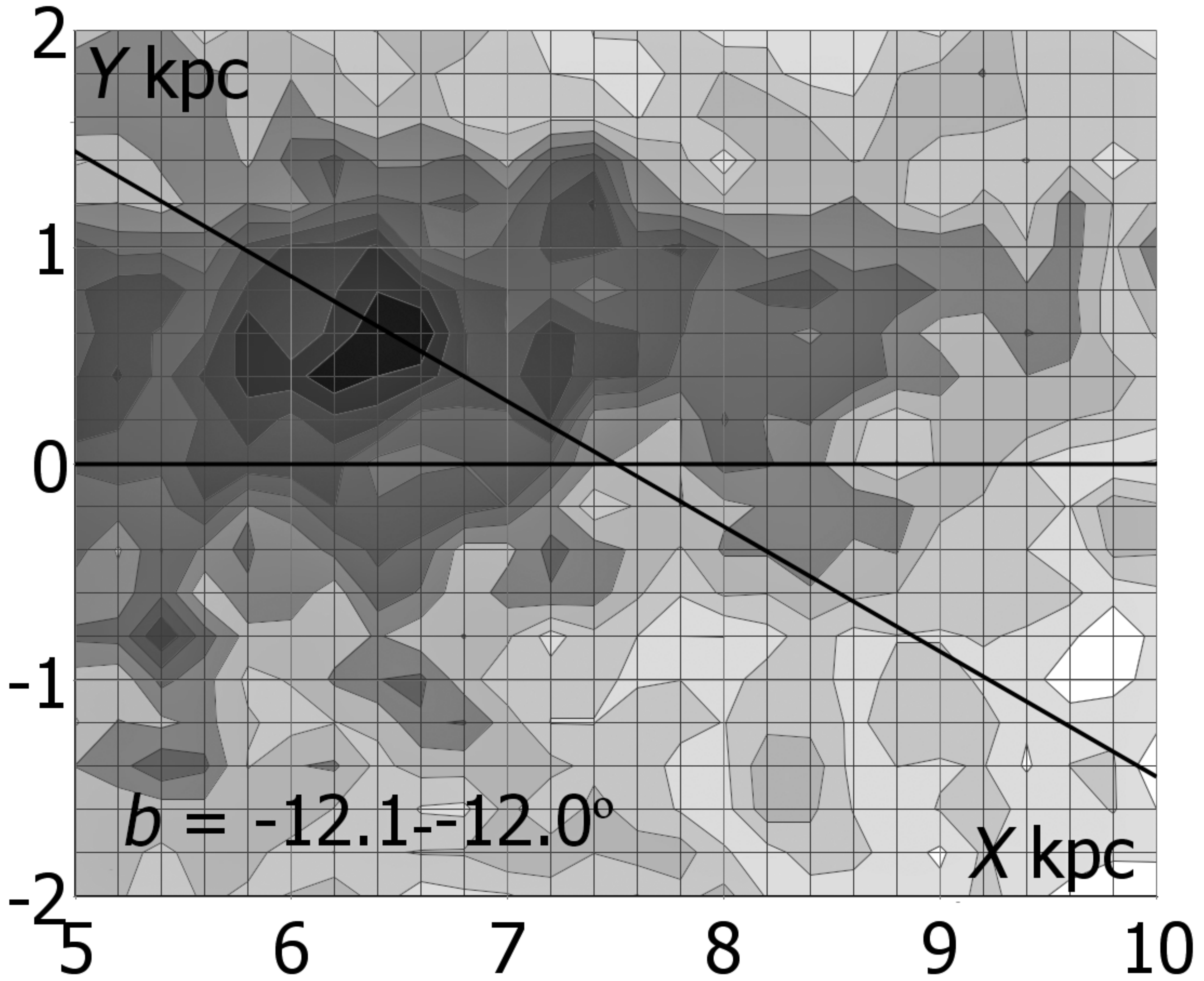}
	\caption{Estimating the distance of the Galactic Centre by triangulation from the densest part of the bar, assuming a bar angle of 30\textdegree. Each plot shows the red clump density for narrow range of latitudes. Peak density lies in the near part of the bar because height above the Galactic plane increases with distance.  }
	\label{Fig:7}
\end{figure}

We restricted colours to the range of the red clump $ 0.5 < J - K \le 0.85 $ mag and plotted the distribution for the central part of the Galaxy using Gaussian smoothing in two dimensions (figure 7) with smoothing parameter 0.1 kpc, which is large enough to remove random noise but small enough to show a the position of the peak to acceptable precision. The solar position is to the left of the plots. Distance errors cause some smearing along the line of sight, but the peak of the distribution is expected at the position of greatest stellar density. 

For slices at fixed latitude, height above the Galactic plane increases with distance from the Sun. It follows that the distribution is expected to peak in the near part of the bar. The distance of the Galactic Centre can be found by drawing a line from the peak of the distribution in the direction of the bar (triangulation, figure 7). Assuming a bar angle of $ 30 \pm 5 $\textdegree{} in agreement with estimates by Francis \& Anderson (2012a), Nataf et al. (2013), Babusiaux and Gilmore (2005) and references in Vanhollebeke et al. (2009) we find $ R_0 = 7.5 \pm 0.2 $ kpc, with good agreement between the six slices used. The error includes a contribution due reddening, assumed to be 20\% of the total reddening correction, which is typically only $ \sim 0.5 $ kpc at these latitudes using the $K$ band. Our result agrees well with the estimates from the halo centroid, with other recalibrated distances from the red clump, and with typical estimates from RR Lyrae and other short period variables. 

From a theoretical study of evolutionary behaviour, Salaris and Girardi (2002) have shown that a population correction should be applied to $ M_K^{ \mathrm{bulge}}(\mathrm{RC}) =M_K^{\mathrm{local}}(\mathrm{RC}) + 0.07 $ mag to take account of differences in age and metallicity between the bulge and the local solar neighbourhood. Nishiyama et al. (2006) estimated the error in the population correction as $ \pm 0.07 $ mag, but suggested that this may be an overestimate. After applying the population correction we find $ R_0 = 7.3 \pm 0.3 $ kpc, to which we add $ \sim 0.2 $ kpc (depending on the precise geometry of the bar) because a fixed latitude means that the peak is expected on the nearer side of the bar. 

\section{Comparison with other measurements}\label{Comparison with other measurements}
$R_0$ kpc was first calculated from the red clump using $I$ band measurements by Paczy\'{n}ski and Stanek (1998), who used Gaussian fits to the luminosity function. Stanek \& Garnavich (1998) made corrections to the treatment and gave the estimate $R_0 =  8.2 \pm 0.15|_{\mathrm{stat}} \pm 0.15|_{\mathrm{sys}} $ kpc, using the calibration $ M_I = -0.23 $ corresponding to the peak of a Gaussian fitted to stars with Hipparcos flag H42 in {A,C,E,F,G}. Using the same calibration, Stanek et al. (2000), gave $ R_0 = 8.67 \pm 0.4 $ kpc, also from red clump stars in Baade's window. 

These estimates are high because the use of flag H42 introduces a selection bias toward bright stars in the calibration sample, because the peaks of the background polynomial plus Gaussian fits used for both bulge and calibration are displaced from the peak of the Gaussian, because population correction effects were not understood before Girardi and Salaris (2001), and because a Gaussian does not well fit a skew distribution and biases the result towards the mean. Table 1 shows a difference of $ \sim 0.2$-$ 0.3 $ magnitudes between the mean and peak magnitudes for the red clump. After taking these effects into account the estimates given by Stanek \& Garnavich (1998), by Stanek et al. (2000) are similar to ours. The estimates of Vanhollebeke et al. (2009) and Nataf et al. (2013) are significantly higher because they use a non-standard extinction law which we have not been able to justify (see below).

Alves (2000) found $ R_0 = 8.24 \pm 0.42 $ kpc using the mean magnitude of a rather small number ($ \sim 20 $) of bulge stars together $ M_K = -1.61 $ mag, based on the original Hipparcos catalogue (Perryman et al. 1997), and using calibrators taken from the same population as Paczy\'{n}ski and Stanek, i.e. using the restriction on the H42 flag. After recalibrating to the mean, $ M_K = -1.424 $, we obtain $ R_0 = 7.6 $ kpc. Babusiaux and Gilmore (2005) and Nishiyama et al (2006) used $ M_K = -1.61 $ from Alves (2000) and $ M_K = -1.60 $ from Alves et al. (2002) respectively, finding $ 7.7 \pm 0.15 $ kpc and $ 7.5 \pm 0.1|_{\mathrm{stat}} \pm 0.35|_{\mathrm{sys}} $ kpc. These estimates already use the peak magnitude for bulge stars, which explains why they are substantially below estimates using a Gaussian or mean fit. Recalibration has less impact and we find 7.3 kpc and 7.2 kpc respectively. McWilliam and Zoccali (2010) found $ 7.3 \pm 0.3 $ kpc from the midpoint of a double peak in the red clump bulge, using $ K $ magnitudes from 2MASS calibrated to red clump stars in 47 Tuc. This calibration is not affected by our study.

We searched the literature for over 150 estimates of the distance to the Galactic Centre (available online). Of these, 137 may be regarded as (broadly) distinct. Prior to 1980 there is a wide scatter of results, indicating that measurements lacked sufficiently precision for a meaningful result, due inadequacies in either theory or in measurement technology. More recent measurements have achieved greater consistency, the lowest since 1980 being 6.8 kpc (Frenk \& White 1982) and the greatest 10.1 kpc (Surdin 1980). Since 2000 the range has narrowed to between 7.2 kpc (B06) and 8.8 kpc (Collinge et al. 2006).

A number of authors (e.g. Malkin 2013, Reid 1993, Nikiforov 2004, Foster \& Cooper 2010) have asked whether measurements of $R_0$ may be subject to a ``bandwagon'' effect, resulting from a reluctance of reviewers or authors to publish figures in poor agreement with preferred estimates. Malkin (2013) suggested that a bandwagon effect might be detected if there is a tendency for figures to gradually approach the true value. Malkin (2013) found no trend in a sample of 52 determinations of $R_0$ published over the last 20 years. Our larger sample shows a continuing small trend toward decreasing estimates, for 48 estimates between 1980 and 2000, the correlation is significant at 92\%, and in 48 determinations since 2000 it remains 73\% significant (by Student's $t$-test). Even if a bandwagon effect is not at work, one might expect a trend towards the true value from an improvement in systematic errors over time. In either case the trend is expected to continue and the true distance to the Galactic Centre is projected to be less than the mean $ R_0 = 8.0 $ kpc found from measurements since 2000. 

Some estimates of luminosity distance have been increased by using non-standard extinction toward the inner Galaxy (Pietrukowicz et al. 2012, Nataf et al. 2013, Vanhollebeke et al. 2009, Collinge et al. 2006). Non-standard selective extinction was proposed by Popowski (2000) to account for the observation that red clump giants and RR Lyrae stars are redder in the bulge and was calculated by Udalski (2003) and Sumi (2004) on the basis that stars in the bulge are the same colour as local stars. However, Kunder et al. (2008, 2010) and McNamara et al. (2000) found no anomalous reddening in studies of RR Lyrae and high amplitude $ \delta $ Scuti, and Girardi and Salaris (2001) have shown that red clump stars in the bulge are expected to be redder than local counterparts, because of the large age and high metallicity of the bulge (e.g. Zoccali et al. 2003). Any anomalous extinction is therefore less than calculated.

As is the case for the red clump, the mean value of $R_0$ from RR Lyrae and short period variables is less for $ K $ band than for $ V $ or $ I $ band measurements: 7.8 kpc from 8 estimates using the $ K $ band and 8.05 kpc from 12 estimates using the $ V $ or $ I $ band. Assuming the validity of the measurements, this also argues against non-standard reddening, which increases $ V $ \& $ I $ band estimates more than $ K $ band estimates. After excluding measurements using non-standard reddening, the mean RR Lyrae distance from the $ V $ and $ I $ bands is 7.9 kpc. These estimates are a little higher than the red clump distance, but agree within a reasonable error. 

An increasingly popular methodology (described by, e.g., Sofue et al. 2011) assumes that objects found to be stationary with respect to the local standard of rest at the Solar position are in circular motion at the Solar radius. It is then possible in principle to calculate the distance to the Galactic Centre. The method has given estimates from 7.2 kpc to 8.4 kpc. However, it requires that exact distances are known and that objects are precisely on the Solar circle. Consequently, good results cannot be expected from individual objects. Sofue et al (2011) and Bobylev (2013) have used data from a number of sources, finding respectively $ R_0 = 7.54 \pm 0.77 $ kpc and a combined estimate $ R_0 = 7.5 \pm 0.3 $ kpc, in good agreement with our estimates.

In recent years much focus has been on attempts to use Keplerian motion of stars close to Sgr A*, particularly S2, since 1992 using the New Technology Telescope and Very Large Telescope at the European Southern Observatory (Eckart \& Genzel 1996), and, since 1995, with the Keck telescope (Ghez et al. 1998). Gillessen et al. (2009) give $R_0 = 8.34 \pm 0.27|_{\mathrm{stat}} \pm 0.52|_{\mathrm{sys}} $ kpc using data on S2 from both teams. The method is extremely sensitive to modelling assumptions so that the tabulated results range over more than 1 kpc, depending upon which assumptions have been used. In view of this sensitivity, we think it problematic that certain factors appear to have been overlooked. One issue with Keplerian motion is that the gravity due to the distribution of matter in the disc and in the bar is not calculated. It is not obvious that it can be ignored (Newton's shell theorem, that gravity due to a spherical shell is zero inside the shell, only applies in approximation in the Galaxy). 

Another issue is that Zucker et al. (2006) have pointed out that Keplerian solutions to S2 orbits do not take account of relativistic effects. Gillessen et al. (2009) commented on de Sitter precession (the geodetic effect), which causes the well known perihelion shift to the orbit of Mercury, finding that it was not possible to measure the pericentre shift of S2 with any accuracy. De Sitter precession arises in Schwarzschild metric. However, if relativistic effects are important then it is necessary to use Kerr metric, since the central black hole is likely to be rotating at close to the relativistic limit. Kerr metric leads to the Lense-Thirring effect, or ``frame-dragging'', made famous and ultimately measured in the vicinity of the Earth by the Gravity Probe B experiment (Everitt et al. 2011). Frame dragging is predicted to affect both the orbit of S2 and the path of the light coming to us from S2, but is not considered by either Zucker et al. (2006) or Gillessen et al. (2009). If present, frame dragging would tend to move S2 from its planar orbit, leading to inaccuracies in the projected orbit and greatly complicating analysis. Bending of the path of light would lead to a further error in position. To our knowledge it has not been shown that this effect is negligible. It therefore appears to us that a complete relativistic analysis of the observations of S2 has not been given.

Both teams reported difficulties concerning the motion of S2 in 2002, when it was near pericentre and relativistic effects were at their greatest. When observations from 2002 are excluded the result is $ 7.72 \pm 0.33 $ kpc (depending on the set of auxiliary assumptions) in reasonable agreement with our estimates. It is also worth noting that the issues described above lead to cumulative differences from Keplerian motion, so that observations over a shorter time scale may actually give a better result. For example, using the same method Eisenhauer et al. (2005) previously found $ 7.6 \pm 0.3 $ kpc. 

\section{Conclusion}\label{Conclusion}
Considerable effort has taken place in the last few years to improve the calibration of the RR Lyrae scale used to determine most cluster distances. Although the scale has been revised upwards, it has not eliminated the difference between the distance from the halo centroid, $R_0 = 7.4 \pm 0.2|_{\mathrm{stat}} \pm 0.2|_{\mathrm{sys}} $ kpc, and the distance found from some other determinations of the Galactic Centre. 

The implications of a greater value of $R_0$ kpc to the RR Lyrae scale, and hence to the cosmological distance scale would be considerable. We calibrated the magnitude of the red-clump to local stars with good parallaxes in Hipparcos, avoiding selection bias in the calibration sample, and calculated the distance to the Galactic Centre from $K$ magnitudes in 2MASS at the periphery of the Bulge. These choices minimise uncertainties due to reddening. We found $ R_0 = 7.5 \pm 0.3 $ kpc in agreement with the distance from the halo centroid. Using our calibration we found agreement with eight other determinations of $R_0$ from the red clump.

Our results agree with $ R_0 = 7.25 \pm 0.32 $ kpc obtained from 19 star-forming regions and with $ R_0 = 7.66 \pm 0.36 $ kpc and $ R_0 = 7.64 \pm 0.32 $ kpc from two populations of Cepheids by Bobylev (2013) assuming circular motions. Our results also agree with $ 7.72 \pm 0.33 $ kpc given by Gillessen et al. (2009) from the orbit of S2 after excluding data from 2002 when S2 was near pericentre; we believe this data should be excluded because of unmodelled effects such as distributed mass in the bar and the relativistic Lense-Thirring effect, or frame dragging, due to the rotation of Sgr A*. Frame dragging is strongest near pericentre. 

For any particular choice of $R_0$, Solar motion must match with the proper motion of Sgr A*, $ 6.379 \pm 0.024 $ mas yr$^{-1}$, measured by Reid and Brunthaler (2004). For $ R_0 = 7.5 $ kpc the component of Solar velocity in the direction of rotation is 227 km s$^{-1}$, on the assumption that Sgr A* is at rest at the Galactic Centre.

\appendix

\onecolumn

\section{Data}
The first ten rows of the data base of globular clusters are shown in table A1 and table A2. The full data base is available online and at CDS, catalogue VII/271.

\begin{table}\label{(table7)}

\begin{tabular}{llrrrrrrrrr} 
ID      	& Name    		& RAdeg   & DEdeg & glon & glat & n  & mu0 & dist & e\_dist & Hdist \\ \\
NGC 104        &47 Tucanae               &  6.0234&-72.0813&305.8949&-44.8894&13& 13.31&  4.58& 0.04&  4.5\\
NGC 288        &                         & 13.1885&-26.5826&151.2851&-89.3804& 4& 14.86&  9.36& 0.19&  8.9\\
NGC 362        &Melotte 4                & 15.8094&-70.8488& 301.533&-46.2474& 6& 14.73&  8.84& 0.13&  8.6\\
Whiting 1      &Whiting 1                & 30.7375& -3.2528&161.6176&-60.6359& 1& 17.34&  29.4&  1.9& 30.1\\
NGC 1261       &                         & 48.0675&-55.2162&270.5387&-52.1244& 4& 16.09& 16.54& 0.37& 16.3\\
Pal 1          &Palomar 1                & 53.3335& 79.5811&130.0648& 19.0281& 2& 15.56& 12.97& 0.59& 11.1\\
AM 1           &E 1                      &   58.76&-49.6067&258.3487&-48.4728& 1& 20.45& 123.2&12.32&123.3\\
Eridanus       &Eridanus star cluster    & 66.1863&  -21.19&218.1103&-41.3324& 2& 19.77& 89.76& 3.72& 90.1\\
Pal 2          &Palomar 2                & 71.5246& 31.3815&170.5302& -9.0722& 1& 17.21& 27.67& 1.46& 27.2\\
NGC 1851       &                         & 78.5286&-40.0461&244.5128&-35.0356& 6&  15.4& 12.01&  0.2& 12.1\\
\end{tabular}

\caption{Cols 1 to 11, first ten rows of globcat.dat. ID    : globular cluster ID -- Name     : alternative names -- RAdeg    : Right ascension (epoch J2000) -- DEdeg    : Declination  (epoch J2000) -- glon     : Galactic longitude -- glat     : Galactic latitude	-- n   : number of sources for distance -- mu0 : distance modulus -- dist     : mean distance		-- e\_dist   : error in distance Hdist    : distance given by H10.}

\end{table}

\begin{table}\label{(table8)}

\begin{tabular}{rrrrrrrrrrrr} 
X& Y  	& Z    & Rgc  &E(B-V)&[Fe/H]& e\_[Fe/H]& r\_[Fe/H]& RV& e\_RV& n\_RV    	& q\_RV \\ \\
  1.9&  -2.6&  -3.2&   6.9&0.04&-0.69& 0.06&S12 &  -17.& 0.2& 9&ti\\
 -0.1&     0&  -9.4&    12&0.03&-1.35& 0.04&S12 & -45.2& 0.4& 7&ti\\
  3.2&  -5.2&  -6.4&   9.2&0.05&-1.31& 0.05&S12 & 222.9& 1.5& 4&ti\\
-13.7&   4.5& -25.6&  33.5&0.03& -0.7&     &H10 &-130.6& 1.8& 1&t \\
  0.1& -10.2& -13.1&  18.1&0.01&-1.28& 0.06&S12 &  63.7&12.1& 2&ti\\
 -7.9&   9.4&   4.2&  18.4&0.15&-0.65& 0.09&H10 & -82.8& 3.3& 1&t \\
-16.5&  -80.& -92.2& 124.4&   0& -1.7& 0.09&H10 &   116&  20& 1&t \\
 -53.& -41.6& -59.3&  94.3&0.02&-1.47& 0.12&S12 &  -21.&   4& 1&t \\
 -27.&   4.5&  -4.4&  34.9&1.24&-1.42& 0.09&H10 &      &  57& 0&  \\
 -4.2&  -8.9&  -6.9&  16.2&0.02&-0.98& 0.11&S12 & 319.4& 0.3& 8&ti\\

\end{tabular}

\caption{Cols 12 to 23, first ten rows of globcat.dat. X  : distance on X-axis	 --  Y  : distance on Y-axis  --  Z   : distance on Z-axis --  Rgc  : distance from G	 --  E(B-V)   : reddening, from H10 --  [Fe/H] : metalicity --  e\_[Fe/H] : error in metalicity --  r\_[Fe/H] : reference for metalicity RV  : radial velocity	 --  e\_RV  : error in radial velocity n\_RV  : number of measurements for RV --  q\_RV  : quality of RV.}

\end{table}

\begin{table}\label{(table9)}
\begin{scriptsize} 
\begin{tabular}{lllllrrr} 
date	&	author	&	bib code	&	method	&	basis	&	$ R_0 $	&	$ \epsilon $	&	use\\
1918.8	&	Shapley	&	1918ApJ....48..154S	&	halo centroid	&	globular clusters	&	13.0	&		&	1\\
1928.8	&	Shapley \& Swope	&	1928PNAS...14..830S	&	bulge luminosity	&	Cepheids in Sco. \& Oph.	&	14.4	&		&	1\\
1930.0	&	Shapley	&	1930PA.....38..628S	&	halo centroid	&	globular clusters	&	16.4	&		&	1\\
1930.2	&	Lindblad	&	1930MNRAS..90..503L	&	Galactic rotation	&	field stars	&	6.5	&		&	1\\
1932.9	&	Stebbins	&	1933PNAS...19..222S	&	halo centroid	&	globular clusters	&	10.0	&		&	1\\
1933.1	&	van de Kamp	&	1933AJ.....42..161V	&	halo centroid	&	globular clusters	&	5.5	&		&	1\\
1934.4	&	Paskett \& Pearce	&	1934MNRAS..94..679P	&	Galactic rotation	&	type O5 to B7 stars	&	10.0	&		&	1\\
1939.1	&	Shapley	&	1939PNAS...25..113S	&	bulge luminosity	&	Cepheids	&	9.7	&		&	1\\
1939.2	&	Joy	&	1939ApJ....89..356J	&	Galactic rotation	&	Cepheids	&	10.0	&		&	1\\
1946.5	&	Mayall	&	1946ApJ...104..290M	&	halo centroid	&	globular clusters	&	9.2	&	1.2	&	1\\
1951.5	&	Baade	&	1951POMic..10....7B	&	bulge luminosity	&	RR Lyrae	&	8.7	&		&	1\\
1953.5	&	Baade	&	1953...............	&	bulge luminosity	&	RR Lyrae	&	8.2	&		&	1\\
1954.4	&	van de Hulst et al.	&	1954BAN....12..117V	&	Galactic rotation	&	H I exterior to $ R_0 $	&	8.3	&		&	1\\
1954.8	&	Weaver	&	1954AJ.....59..375W	&	circular motion tracers	&	Cepheids in open clusters	&	8.8	&		&	1\\
1958.0	&	Feast \& Thackeray	&	1958MNRAS.118..125F	&	Galactic rotation	&	B type stars	&	8.9	&		&	1\\
1960.0	&	Kron \& Mayall	&	1960AJ.....65..581K	&	bulge luminosity	&	RR Lyrae	&	12.0	&	1.2	&	1\\
1960.0	&	Kron \& Mayall	&	1960AJ.....65..581K	&	halo centroid	&	globular clusters	&	12.5	&	1.5	&	1\\
1961.8	&	Brandt	&	1961PASP...73..324B	&	Galactic mass estimate	&	analogy with Andromeda	&	10.5	&	1.5	&	1\\
1962.7	&	Fernie	&	1962AJ.....67..769F	&	halo centroid	&	globular clusters	&	9.3	&		&	1\\
1963.2	&	Takase	&	1963AJ.....68...80T	&	Galactic rotation	&	Cepheids	&	11.1	&		&	1\\
1965.5	&	Schmidt	&	1965gast.conf..513S	&	Galactic rotation	&	Oort Constants	&	10.0	&		&	1\\
1965.5	&	Arp	&	1965gast.conf..401A	&	halo centroid	&	globular clusters	&	9.9	&	0.5	&	1\\
1965.6	&	Feast \& Shuttleworth	&	1965MNRAS.130..245F	&	Galactic rotation	&	B type stars	&	9.9	&	0.9	&	1\\
1967.0	&	Feast	&	1967MNRAS.136..141F	&	Galactic rotation	&	Cepheids	&	9.8	&	1.4	&	1\\
1970.3	&	Takase	&	1970PASJ...22..255T	&	Galactic rotation	&	Cepheids	&	10.0	&		&	1\\
1972.5	&	Hartwick et al.	&	1972ApJ...174..573H	&	bulge luminosity	&	RR Lyrae	&	7.0	&	0.6	&	1\\
1972.5	&	Toomre	&	1972QJRAS..13..241T	&	Galactic rotation	&	H I interior to $ R_0 $	&	8.3	&	0.5	&	1\\
1973.2	&	Plaut	&	1973A\&A....26..317P	&	bulge luminosity	&	RR Lyrae	&	8.3	&		&	1\\
1973.8	&	Balona \& Feast	&	1974MNRAS.167..621B	&	Galactic rotation	&	OB stars	&	9.0	&	1.6	&	1\\
1974.1	&	van den Bergh \& Herbst	&	1974AJ.....79..603V	&	bulge luminosity	&	RR Lyrae	&	9.2	&	2.2	&	1\\
1974.2	&	Cruz-Gonzalez	&	1974MNRAS.168...41C	&	Galactic rotation	&	nearby stars	&	8.9	&	0.5	&	1\\
1974.7	&	Rybicki et al.	&	1974BAAS....6..453R	&	Galactic rotation	&	Oort Constants	&	9.0	&		&	1\\
1975.2	&	Oort \& Plaut	&	1975A\&A....41...71O	&	bulge luminosity	&	RR Lyrae	&	8.7	&	0.6	&	1\\
1976.1	&	Harris	&	1976AJ.....81.1095H	&	halo centroid	&	globular clusters	&	8.5	&	1.6	&	1\\
1976.7	&	Crampton et al.	&	1976MNRAS.176..683C	&	circular motion tracers	&	OB stars	&	8.2	&		&	1\\
1977.0	&	Sasaki \& Ishizawa	&	1978A\&A....69..381S	&	halo centroid	&	globular clusters	&	9.2	&	1.3	&	1\\
1977.5	&	Belikov \& Syrovoj	&	1977ATsir.968....5B	&	halo centroid	&	globular clusters	&	8.5	&		&	1\\
1977.6	&	Byl \& Ovenden	&	1978ApJ...225..496B	&	Galactic rotation	&	OB stars, Cepheids, open clusters	&	10.4	&	0.5	&	1\\
1978.1	&	Loktin	&	1979SvA....23..671L	&	Galactic rotation	&	OB stars	&	8.1	&	1.7	&	1\\
1978.5	&	de Vaucouleurs \& Buta	&	1978AJ.....83.1383D	&	halo centroid	&	globular clusters	&	7.0	&		&	1\\
1979.4	&	Clube \& Dawe	&	1980MNRAS.190..591C	&	Galactic rotation	&	RR Lyrae \& Cepheids	&	7.0	&	0.1	&	1\\
1979.7	&	Harris	&	1980IAUS...85...81H	&	halo centroid	&	globular clusters	&	8.0	&	1.4	&	0\\
1979.7	&	Harris	&	1980IAUS...85...81H	&	halo centroid	&	globular clusters	&	8.5	&	1.0	&	0\\
1980.1	&	Quiroga	\&	1980A&A....92..186Q	&	Galactic rotation	&	H I, OB stars, H II regions	&	8.5	&	0.7	&	1\\
1980.6	&	Wilson \& Raymond	&	1930AJ.....40..121R	&	Galactic rotation	&	field stars	&	8.8	&		&	1\\
1980.8	&	Surdin	&	1980SvA....24..550S	&	halo centroid	&	globular cluster metallicity	&	10.1	&		&	1\\
1980.8	&	Caldwell \& Ostriker	&	1981ApJ...251...61C	&	Galactic rotation	&	Galactic mass model	&	9.1	&	0.6	&	1\\
1981.2	&	Glass \& Feast	&	1982MNRAS.198..199G	&	bulge luminosity	&	Miras	&	9.2	&	0.6	&	1\\
1981.2	&	Glass \& Feast	&	1982MNRAS.198..199G	&	bulge luminosity	&	Miras	&	8.8	&		&	0\\
1982.0	&	Frenk \& White	&	1982MNRAS.198..173F	&	halo centroid	&	globular clusters	&	6.8	&	0.8	&	1\\
1983.4	&	Ostriker	&	1983ASSL..100..249O	&	Galactic rotation	&	Galactic mass model	&	8.2	&		&	1\\
1983.7	&	Herman	&	1983PhDT.......130H	&	Galactic rotation	&	masers	&	9.2	&	1.2	&	1\\
1983.7	&	Ebisuzaki	&	1984PASJ...36..551E	&	halo centroid	&	X-Ray Bursts	&	7.0	&		&	1\\
1985.3	&	Rohlfs et al.	&	1986A&A...158..181R	&	Galactic rotation	\&	H II Regions	&	7.9	&	0.7	&	1\\
1985.5	&	Walker \& Mack	&	1986MNRAS.220...69W	&	bulge luminosity	&	RR Lyrae	&	8.1	&	0.4	&	1\\
1985.7	&	Iurevich	&	1985Afz....23..265I	&	Galactic rotation	&	hydroxyl clouds	&	8.2	&	0.1	&	1\\
1985.7	&	Blanco \& Blanco	&	1985MmSAI..56...15B	&	bulge luminosity	&	RR Lyrae	&	7.3	&	0.5	&	1\\
1985.7	&	Blanco \& Blanco	&	1985MmSAI..56...15B	&	bulge luminosity	&	RR Lyrae	&	8.0	&	0.7	&	0\\
1985.7	&	Blanco \& Blanco	&	1985MmSAI..56...15B	&	bulge luminosity	&	RR Lyrae	&	6.9	&	0.6	&	0\\
1986.3	&	Fernley et al.	&	1987MNRAS.226..927F	&	bulge luminosity	&	RR Lyrae	&	8.0	&	0.7	&	1\\
1987.0	&	Caldwell \& Coulson	&	1987AJ.....93.1090C	&	Galactic rotation	&	Cepheids	&	7.8	&	0.7	&	1\\
1987.4	&	Blitz \& Brand	&	1988LNP...306...73B	&	circular motion tracers	&	H II Regions	&	8.0	&	0.5	&	1\\
1988.5	&	Reid et al.	&	1988ApJ...330..809R	&	centre parallax	&	masers	&	7.1	&	1.5	&	1\\
1989.2	&	Racine \& Harris	&	1989AJ.....98.1609R	&	halo centroid	&	globular clusters	&	7.5	&	0.9	&	1\\
1990.7	&	Pottasch	&	1990A\&A...236..231P	&	bulge luminosity	&	planetary nebulae	&	7.8	&	0.8	&	1\\
1991.5	&	Dopita et al.	&	1992ApJ...389...27D	&	bulge luminosity	&	planetary nebulae	&	7.6	&	0.7	&	1\\
1991.7	&	Walker \& Terndrup	&	1991ApJ...378..119W	&	bulge luminosity	&	RR Lyrae	&	8.1	&	0.5	&	1\\
1991.7	&	Gwinn et al.	&	1992ApJ...393..149G	&	Galactic rotation	&	masers	&	8.1	&	1.1	&	1\\
1991.7	&	Merrifield	&	1992AJ....103.1552M	&	Galactic rotation	&	H I	&	7.9	&	0.8	&	1\\
1992.1	&	Whitelock	&	1992ASPC...30...11W	&	bulge luminosity	&	Miras	&	9.1	&		&	1\\
1992.2	&	Moran et al.	&	1993LNP...412..244M	&	circular motion tracers	&	masers	&	7.6	&	0.6	&	1\\
1993.4	&	Maciel	&	1993Ap\&SS.206..285M	&	halo centroid	&	globular clusters	&	7.6	&	0.4	&	1\\
1993.5	&	Moran	&	1993AAS...182.2704M	&	circular motion tracers	&	masers	&	7.8	&	0.6	&	1\\
1994.3	&	Pont et al.	&	1994A\&A...285..415P	&	Galactic rotation	&	Cepheids	&	8.1	&	0.3	&	1\\
1994.4	&	Rastorguev et al.	&	1994AstL...20..591R	&	halo centroid	&	globular clusters	&	7.0	&	0.5	&	1\\
1994.5	&	Nikiforov	&	1994...............	&	halo centroid	&	globular cluster kinetics	&	7.0	&		&	1\\
1994.7	&	Nikiforov \& Petrovskaya	&	1994ARep...38..642N	&	Galactic rotation	&	H I \& H II Data	&	7.5	&	1.0	&	1\\
1994.8	&	Glass et al.	&	1995MNRAS.273..383G	&	bulge luminosity	&	Miras	&	8.7	&	0.7	&	1\\
1994.9	&	Dambis et al.	&	1995AstL...21..291D	&	Galactic rotation	&	Cepheids	&	7.1	&	0.5	&	1\\
1995.3	&	Carney et al.	&	1995AJ....110.1674C	&	bulge luminosity	&	RR Lyrae	&	7.8	&	0.4	&	1\\
1995.6	&	Huterer et al.	&	1995AJ....110.2705H	&	dynamical solution	&	M type giants	&	8.2	&	1.0	&	1\\
1996.4	&	Feast	&	1997MNRAS.284..761F	&	bulge luminosity	&	RR Lyrae	&	8.1	&	0.4	&	1\\
\end{tabular}
\caption{Time line of estimates of distance to the Galactic centre. Error includes systematic error in quadrature, when quoted.}
\end{scriptsize}
\end{table}
\begin{table}\label{(table9)}
\begin{scriptsize}
\begin{tabular}{lllllrrr}
1996.5	&	Honma \& Sofue	&	1996PASJ...48L.103H	&	Galactic rotation	&	OH/IR stars \& young stars	&	7.6	&		&	1\\
1996.6	&	Layden et al.	&	1996AJ....112.2110L	&	bulge luminosity	&	RR Lyrae	&	7.6	&	0.4	&	1\\
1997.5	&	Feast \& Whitelock	&	1997MNRAS.291..683F	&	Galactic rotation	&	Cepheids	&	8.5	&	0.5	&	1\\
1997.6	&	Paczynski \& Stanek	&	1998ApJ...494L.219P	&	bulge luminosity	&	red clump	&	8.4	&	0.4	&	0\\
1998.1	&	Olling \& Merrifield	&	1998MNRAS.297..943O	&	Galactic rotation	&	metastudy	&	7.1	&	0.4	&	1\\
1998.1	&	Stanek \& Garnavich	&	1998ApJ...503L.131S	&	bulge luminosity	&	red clump	&	8.2	&	0.2	&	1\\
1998.1	&	Udalski	&	1998AcA....48..113U	&	bulge luminosity	&	RR Lyrae	&	8.1	&	0.2	&	1\\
1998.3	&	Glushkova et al.	&	1998A\&A...329..514G	&	Galactic rotation	&	pop. I objects	&	7.3	&	0.3	&	1\\
1998.3	&	Surdin	&	1999A\&AT...18..367S	&	halo centroid	&	globular cluster metallicity	&	8.6	&	1.0	&	1\\
1998.3	&	Morgan et al.	&	1998AcA....48..509M	&	bulge luminosity	&	Delta Scuti	&	7.6	&		&	1\\
1998.4	&	Feast et al.	&	1998MNRAS.298L..43F	&	Galactic rotation	&	Cepheids	&	8.5	&	0.3	&	1\\
1998.7	&	Catchpole et al.	&	1999IAUS..192...89C	&	bulge luminosity	&	Miras	&	9.4	&	0.5	&	1\\
1998.8	&	Metzger et al.	&	1998AJ....115..635M	&	Galactic rotation	&	Cepheids	&	7.7	&	0.3	&	1\\
2000.1	&	Genzel et al.	&	2000MNRAS.317..348G	&	dynamical solution	&	Galactic centre stars	&	7.9	&	0.9	&	1\\
2000.1	&	McNamara et al.	&	2000PASP..112..202M	&	bulge luminosity	&	Delta Scuti \& RR Lyrae	&	7.9	&	0.3	&	1\\
2000.2	&	Alves	&	2000ApJ...539..732A	&	bulge luminosity	&	red clump	&	8.2	&	0.4	&	1\\
2000.5	&	Nikiforov	&	2000ASPC..209..403N	&	Galactic rotation	&	H I regions	&	8.2	&	0.7	&	1\\
2000.6	&	Stanek et al.	&	2000AcA....50..191S	&	bulge luminosity	&	red clump	&	8.7	&	0.4	&	1\\
2003.4	&	Eisenhauer et al.	&	2003ApJ...597L.121E	&	dynamical solution	&	central star cluster	&	7.2	&	0.9	&	1\\
2003.4	&	Eisenhauer et al.	&	2003ApJ...597L.121E	&	dynamical solution	&	S2	&	8.0	&	0.4	&	0\\
2004.1	&	Gerasimenko	&	2004ARep...48..103G	&	Galactic rotation	&	open clusters	&	8.3	&	0.3	&	1\\
2005.0	&	Babusiaux \& Gilmore	&	2005MNRAS.358.1309B	&	bulge luminosity	&	red clump	&	7.7	&	0.2	&	1\\
2005.0	&	Babusiaux \& Gilmore	&	2005MNRAS.358.1309B	&	bulge luminosity	&	red clump	&	7.6	&	0.2	&	0\\
2005.1	&	Eisenhauer et al.	&	2005ApJ...628..246E	&	dynamical solution	&	S2	&	7.6	&	0.3	&	0\\
2005.4	&	Avedisova	&	2005ARep...49..435A	&	Galactic rotation	&	mol. gas in star forming regions	&	8.0	&	0.4	&	1\\
2005.5	&	Groenewegen \& Blommaert	&	2005A\&A...443..143G	&	bulge luminosity	&	Miras	&	8.7	&	0.4	&	1\\
2005.5	&	Groenewegen \& Blommaert	&	2005A\&A...443..143G	&	bulge luminosity	&	Miras	&	8.6	&	0.7	&	0\\
2005.5	&	Groenewegen \& Blommaert	&	2005A\&A...443..143G	&	bulge luminosity	&	Miras	&	8.8	&	0.4	&	0\\
2005.7	&	Zucker	&	2006ApJ...639L..21Z	&	dynamical solution	&	S2	&	7.7	&	0.3	&	0\\
2005.9	&	Bica et al.	&	2006A\&A...450..105B	&	halo centroid	&	globular clusters	&	7.2	&	0.3	&	1\\
2006.0	&	Collinge et al.	&	2006ApJ...651..197C	&	bulge luminosity	&	RR Lyrae	&	8.8	&	0.7	&	1\\
2006.5	&	Nishiyama et al.	&	2006ApJ...647.1093N	&	bulge luminosity	&	red clump	&	7.5	&	0.4	&	1\\
2007.1	&	Shen \& Zhu	&	2007ChJAA...7..120S	&	Galactic rotation	&	(combined)	&	8.1	&	0.5	&	0\\
2007.1	&	Shen \& Zhu	&	2007ChJAA...7..120S	&	Galactic rotation	&	OB stars	&	8.3	&	0.8	&	1\\
2007.1	&	Shen \& Zhu	&	2007ChJAA...7..120S	&	Galactic rotation	&	open clusters	&	8.0	&	0.6	&	1\\
2007.7	&	Bobylev et al.	&	2007AstL...33..720B	&	Galactic rotation	&	open clusters	&	7.4	&	0.3	&	1\\
2008.0	&	Groenewegen et al.	&	2008A\&A...481..441G	&	bulge luminosity	&	Cepheids \& RR Lyrae	&	7.9	&	0.5	&	1\\
2008.1	&	Feast et al.	&	2008MNRAS.386.2115F	&	bulge luminosity	&	Cepheids \& RR Lyrae	&	7.6	&	0.2	&	1\\
2008.1	&	Feast et al.	&	2008MNRAS.386.2115F	&	bulge luminosity	&	Cepheids \& RR Lyrae	&	8.2	&	0.6	&	0\\
2008.4	&	Kunder \& Chaboyer	&	2008AJ....136.2441K	&	bulge luminosity	&	RR Lyrae	&	8.0	&	0.3	&	1\\
2008.6	&	Ghez et al.	&	2008ApJ...689.1044G	&	dynamical solution	&	S2	&	8.0	&	0.6	&	1\\
2008.6	&	Ghez et al.	&	2008ApJ...689.1044G	&	dynamical solution	&	S2	&	8.4	&	0.4	&	0\\
2008.8	&	Trippe et al.	&	2008A\&A...492..419T	&	dynamical solution	&	Galactic centre stars	&	8.1	&	0.3	&	1\\
2008.8	&	Gillessen et al.	&	2009ApJ...692.1075G	&	dynamical solution	&	S2	&	8.3	&	0.4	&	1\\
2009.2	&	Vanhollebeke et al.	&	2009A\&A...498...95V	&	bulge luminosity	&	red clump	&	8.8	&	0.5	&	1\\
2009.3	&	Reid et al.	&	2009ApJ...700..137R	&	circular motion tracers	&	masers	&	8.4	&	0.6	&	1\\
2009.4	&	Majaess et al.	&	2009MNRAS.398..263M	&	bulge luminosity	&	Cepheids	&	7.8	&	0.6	&	1\\
2009.4	&	Majaess et al.	&	2009MNRAS.398..263M	&	bulge luminosity	&	Cepheids	&	7.7	&	0.7	&	0\\
2009.4	&	Majaess et al.	&	2009MNRAS.398..263M	&	bulge luminosity	&	Cepheids	&	7.8	&	0.6	&	0\\
2009.4	&	Dambis	&	2009MNRAS.396..553D	&	bulge luminosity	&	RR Lyrae	&	7.6	&	0.4	&	1\\
2009.5	&	Matsunaga et al.	&	2009MNRAS.399.1709M	&	bulge luminosity	&	Miras	&	8.2	&	0.4	&	1\\
2009.6	&	Reid et al.	&	2009ApJ...705.1548R	&	centre parallax	&	masers	&	7.9	&	0.8	&	1\\
2009.8	&	Gillessen et al.	&	2009ApJ...707L.114G	&	dynamical solution	&	S2	&	8.3	&	0.3	&	1\\
2010.0	&	Dambis	&	2010arXiv1001.1428D	&	bulge luminosity	&	RR Lyrae	&	7.7	&	0.4	&	1\\
2010.1	&	Majaess	&	2010AcA....60...55M	&	bulge luminosity	&	RR Lyrae	&	8.1	&	0.6	&	1\\
2010.5	&	Sato et al.	&	2010ApJ...720.1055S	&	circular motion tracers	&	masers	&	8.3	&	1.2	&	1\\
2010.7	&	McWilliam \& Zoccali	&	2010ApJ...724.1491M	&	bulge luminosity	&	red clump	&	7.3	&	0.3	&	1\\
2011.0	&	Ando et al.	&	2011PASJ...63...45A	&	circular motion tracers	&	masers	&	7.8	&	0.3	&	1\\
2011.1	&	McMillan	&	2011MNRAS.414.2446M	&	Galactic rotation	&	various	&	8.3	&	0.2	&	1\\
2011.3	&	Liu \& Zhu	&	2011ASPC..451..339L	&	Galactic rotation	&	open clusters	&	8.0	&	0.7	&	1\\
2011.3	&	Sofue et al.	&	2011PASJ...63..867S	&	circular motion tracers	&	H II Regions \& masers	&	7.5	&	0.8	&	1\\
2011.5	&	Pietrukowicz et al.	&	2012ApJ...750..169P	&	bulge luminosity	&	RR Lyrae	&	8.5	&	0.4	&	1\\
2012.5	&	Schoenrich	&	2012MNRAS.427..274S	&	Galactic rotation	&	SEGUE stars	&	8.3	&	0.3	&	1\\
2012.8	&	Nataf et al.	&	2013ApJ...769...88N	&	bulge luminosity	&	red clump	&	8.2	&	0.4	& 1	\\
2012.8	&	Matsunaga et al.	&	2013MNRAS.429..385M	&	bulge luminosity	&	short-period variables	&	8.1	&	0.4	&	1\\
2012.9	&	Honma et al.	&	2012PASJ...64..136H	&	Galactic rotation	&	masers	&	8.1	&	0.5	&	1\\
2013.0	&	Bobylev	&	2013AstL...39...95B	&	circular motion tracers	&	(combined)	&	7.5	&	0.3	&	0\\
2013.0	&	Bobylev	&	2013AstL...39...95B	&	circular motion tracers	&	Cepheids 	&	7.6	&	0.3	&	1\\
2013.0	&	Bobylev	&	2013AstL...39...95B	&	circular motion tracers	&	Cepheids 	&	7.7	&	0.4	&	1\\
2013.0	&	Bobylev	&	2013AstL...39...95B	&	circular motion tracers	&	star forming regions 	&	7.3	&	0.3	&	1\\
2013.9	&	Francis \& Anderson	&	This paper	&	bulge luminosity	&	red clump	&	7.5	&	0.3	&	1\\
2013.9	&	Francis \& Anderson	&	This paper	&	halo centroid	&	globular clusters	&	7.4	&	0.3	&	1

\end{tabular}
\end{scriptsize}

\caption{Time line of estimates of distance to the Galactic centre. Error includes systematic error in quadrature, when quoted.}

\end{table}

\twocolumn

\label{lastpage}

\end{document}